\documentclass[trackchanges,twocolumn]{aastex61}

\usepackage{amsmath, amsthm, amssymb}
\usepackage{bm}
\usepackage{multirow}
\usepackage{xcolor}                                                                      
\usepackage{colortbl} 
\accepted{\today}
\submitjournal{ApJS}

\shorttitle{The MRI in Mesh-Free Methods}
\shortauthors{Deng et al.}

\begin{document}

\title{Local simulations of MRI turbulence with meshless methods}

\correspondingauthor{Hongping Deng}
\email{hpdeng@physik.uzh.ch}

\author{Hongping Deng}
\affiliation{Center for Theoretical Astrophysics and Cosmology, Institute for Computational Science, University of Zurich, Winterthurerstrasse 190, 8057 Zurich, Switzerland}

\author{Lucio Mayer}
\affiliation{Center for Theoretical Astrophysics and Cosmology, Institute for Computational Science, University of Zurich, Winterthurerstrasse 190, 8057 Zurich, Switzerland}

\author{Henrik Latter}
\affiliation{Department of Applied Mathematics and Theoretical Physics, University of Cambridge, Centre for Mathematical Sciences, Wilberforce Road, Cambridge
CB3 0WA, UK}

\author{Philip F. Hopkins}
\affiliation{TAPIR, California Institute of Technology, Pasadena, CA 91125, USA}

\author{Xue-Ning Bai}
\affiliation{Institute for Advanced Study and Tsinghua Center for Astrophysics, Tsinghua University, Beijing 100084, China}

\begin{abstract}
  The magneto-rotational instability (MRI) is one of the most important processes in sufficiently ionized astrophysical disks. Grid-based simulations, especially those using the local shearing box approximation, provide a powerful tool to study the nonlinear turbulence the MRI produces. On the other hand, meshless methods have been widely used in cosmology, galactic dynamics, and planet formation, but have not been fully deployed on the MRI problem. We present local unstratified and vertically stratified MRI simulations with two meshless MHD schemes: a recent implementation of SPH MHD \citep{Price2012}, and a MFM MHD scheme with constrained gradient divergence cleaning, as implemented in the GIZMO code \citep{Hopkins2017}. Concerning variants of the SPH hydro force formulation, we consider both the ``vanilla" SPH and the PSPH variant included in GIZMO.
  We find, as expected, that the numerical noise inherent in these schemes affects  turbulence significantly. Also a high-order kernel, free of the pairing instability, is necessary. Both schemes adequately simulate MRI turbulence in unstratified shearing boxes with net vertical flux. The turbulence, however, dies out in zero-net-flux unstratified boxes, probably due to excessive numerical dissipation. In zero-net-flux vertically stratified simulations, MFM can reproduce the MRI dynamo and its characteristic butterfly diagram for several tens of orbits before ultimately decaying. In contrast, extremely strong toroidal fields, as opposed
  to sustained turbulence, develop in equivalent simulations using SPH MHD. The latter unphysical state is likely caused by a combination of excessive artificial viscosity, numerical resistivity, and the relatively large residual errors in the
  divergence of the magnetic field.
\end{abstract}




\section{Introduction} 
\label{sec:intro}
The turbulence instigated by the magneto-rotational instability (MRI) can transport angular momentum outwards thus enabling accretion in several sources, such as dwarf novae, low mass X-ray binaries, and Active Galactic Nuclei (AGNs). Numerical MHD simulations are necessary to study this highly non-linear problem. Simulations of MRI range from local shearing box simulations, unstratified \citep[e.g.][]{Hawley1995, Hawley1996, Sano2004,Simon2009} and stratified \citep[e.g.][]{Brandenburg1995, Stone1996, Miller2000, Davis2010, Simon2011}, to global simulations \citep[e.g.][]{Armitage1998, Hawley2000,Steinacker2002,Fromang2006, Parkin2013, Zhu2018}. Three-dimensional simulations carried out with different grid-based codes, such as ZEUS \citep{Hawley1995}, Pencil \citep{Brandenburg2002}, RAMSES \citep{Teyssier2002, Fromang2006}, ATHENA \citep{Stone2008}, and the spectral code Snoopy \citep{Lesur2007} report similar statistics for the turbulence. 

Local MRI simulations are especially challenging because the saturated state appears to depend on the small-scale diffusion, be it physical or numerical. For instance, zero-net-flux simulations in unstratified boxes do not converge with increasing resolution, as the turbulent motions reach their peak amplitude near the smallest resolvable scales. But if physical sources of diffusivity are incorporated and resolved, turbulence can die out when the magnetic Prandtl number is too small \citep{Fromang2007a, Fromang2007b}. The latter dissipation is also sensitive to the vertical aspect ratio of the computational domain  ($L_z/L_x$) \citep{Shi2015}. On the other hand, in net vertical flux simulations angular momentum transport depends on the magnetic Prandtl number yet again, at least when the latter takes values of order unity \citep{Meheut2015}. Vertically stratified shearing box simulations without a net flux also suffer convergence problems \citep{Bodo2014,Ryan2017}. Adding a net vertical flux, however, can radically change the character of MRI turbulence as, for example, magnetic winds, may be launched. Some of the properties of these winds also suffer from non-convergence \citep{Fromang2013, Bai2013, Lesur2013}. Finally, non-ideal MHD effects can suppress or radically alter the nature and strength of turbulence \citep[see, e.g.][]{Fleming2000, Turner2007, Bai2011, Lesur2014, Bai2014, Simon2015}. Such effects are still under investigation. 

Currently there are very few published  studies of the MRI undertaken with mesh-free methods \citep[see,e.g.,][]{Gaburov2011,Pakmor2013,Hopkins2015b}, and no systematic exploration of MRI properties over the various standard flow and magnetic field 
configurations routinely examined
with grid-based codes. This is despite 
the frequent use of smoothed-particle magnetohydrodynamics (SPH MHD) \citep{Springel2010, Price2012}, in particular, to probe galaxy, star, and planet formation \citep{Dobbs2016, Dolag2009, Steinwandel2018,Price2007, Price2008,Price2009,Forgan2016}. The neglect is perhaps connected to several numerical problems specific to meshless methods, which we now briefly discuss. 

In codes without meshes or with arbitrary mesh geometries, minimizing the divergence of magnetic fields \citep{Tricco2012, Hopkins2016a} is a major challenge. Because proper minimisation of the divergence is hard to achieve, small ``magnetic monopoles'' can arise, leading to spurious magnetic field reconfiguration, re-connection, and artificial dissipation in 
neighboring domains. 
On fixed, rectilinear, regular, non-moving grids, the \emph{Constrained Transport} (CT) scheme \citep{Evans1988} can maintain zero divergence to machine precision.  Until recently, CT schemes had only been implemented for regular, non-moving meshes, but recently \citet{Mocz2014,Mocz2016} successfully generalized the CT method to moving meshes that adopt a Voronoi tesselation as their volume partition (e.g.\ those in AREPO, \citealt{Springel2010a}). However, most Lagrangian or quasi-Lagrangian methods, including  moving-mesh as well as particle-based methods or mesh-free finite-volume methods, use ``divergence cleaning'' schemes to keep $\bm{\nabla}\cdot \bm{B}$ minimal \citep{Powell1999, Dedner2002}. \citet{Tricco2012} developed improved divergence-cleaning implementations in SPH (adapting the hyperbolic cleaning scheme from \citealt{Dedner2002}), and showed this could successfully reproduce some standard MHD tests, for example the Orszag-Tang vortex. But in non-linear MRI simulations, and in fact in any regime of  MHD turbulence, effective divergence cleaning is especially difficult due to the complex, multi-scale field geometry,
hence the latter methods are not guaranteed to work satisfactorily.

SPH also suffers from known numerical dissipation sourced by various terms, including the E0 error \citep{Read2010},  pairing instability \citep{Rosswog2015, Dehnen2012} and incorrectly-triggered artificial viscosity \citep{Deng2017a}. It is well-known that this additional numerical dissipation impedes SPH's capability to model subsonic turbulence, even {\em without}  magnetic fields \citep{Bauer2012, Hopkins2015a, Deng2017b}.

The Lagrangian mesh-less finite-volume (MFV) method was developed two decades ago \citep[see, e.g.,][]{Vila1999,Hietel2000} 
and has been significantly improved since \citep[see, e.g.][]{Lanson2008a, Lanson2008b}. 
Recently it has stimulated a growing interest in the astrophysical community \citep[see, e.g.,][]{Gaburov2011,Hubber2017}. 
\citet{Hopkins2015a} generalized the method in \citet{Gaburov2011} to other mesh-free finite-volume Godunov schemes, 
including the closely-related ``meshless finite-mass'' (MFM) method.
  These methods, similarly to moving mesh methods, attempt to combine advantages of grid-based and particle-based  codes. 
In particular, they can describe subsonic hydrodynamical turbulence reasonably well 
(with a quality comparable to regular mesh grid-codes; \citealt{Hopkins2015a}), 
though at greater computational cost; they avoid advection problems in complex flow geometries that are better
modeled in the Lagrangian frame, respecting, for example, Galilean invariance,  and finally they can 
naturally extend to self-gravitating flows  by exploiting accurate state-of-the-art gravity solvers, such as treecodes,
which have native implementations in particle codes. 
Mesh-less methods have been generalized to MHD \citep{Hopkins2016a}, and in subsequent work by \citet{Hopkins2016b} a constrained-gradient (CG) divergence cleaning scheme  has been developed that can maintain a much smaller $\bm{\nabla} \cdot\bm{B}$  (by $\sim$2 orders of magnitude) compared to hyperbolic divergence cleaning. These methods, as implemented in the public code GIZMO,\footnote{The public version of the code, containing all the algorithms used here, is available at \href{http://www.tapir.caltech.edu/~phopkins/Site/GIZMO.html}{\url{http://www.tapir.caltech.edu/~phopkins/Site/GIZMO.html}}} have in fact already been used to simulate the MRI in two-dimensional unstratified shearing sheets \citep{Hopkins2015b}, and these tests have demonstrated that it recovers the correct linear growth rates and behaves similarly to well-tested grid codes (e.g.\ ATHENA). However, how these methods perform in three dimensions, in stratified
configurations, and/or during non-linear saturation, remain untested.

In this paper, we carried out MRI simulations in both  unstratified and vertically stratified shearing boxes, with both SPH and MFM MHD implementations as they are implemented in the multi-method GIZMO code, in order to explore the numerical requirements for these methods to treat the MRI in the non-linear regime. We focus on MFM as opposed to MFV or more general moving-mesh schemes (several of which are also implemented in GIZMO and can, in principle, use the same constrained-gradient divergence ``cleaning'' method) because MFM is designed, such as SPH, to conserve exactly the mass of fluid elements (i.e.\ there is identically zero advection), so the method is ``purely'' Lagrangian. This is perhaps the most challenging case for our purpose, since hybrid moving-mesh or MFV-type methods, in which the grid moves but mass fluxes are also allowed, effectively 
act as a smoothing of grid motion, thus interpolating 
between the ``pure Lagrangian'' (constant mesh-motion) and the ``pure Eulerian'' (fixed-grid) representation of a fluid. 

We will explore both the traditional ``density-energy'' formulation of SPH (named hereafter `TSPH') \citep{Springel2005} and the more recently-developed ``pressure-energy'' formulation (`PSPH') \citep{Saitoh2013,Hopkins2012}. The SPH MHD used here represents the state-of-the-art implementation described by \citet{Price2012} with the advanced artificial viscosity/resistivity switches developed in \citep{Cullen2010, Tricco2013} and divergence cleaning following \citep{Tricco2012}. In unstratified shearing box simulations, no significant density contrast is present and we expect TSPH and PSPH to perform similarly. Therefore, we did not run TSPH and PSPH comparisons for this particular setup. In the GIZMO MFM runs we adopt the constrained-gradient divergence cleaning of \citet{Hopkins2016b}.

We note that the effect of the initial noise in MFV, which appears  to depend on the regularity of the initial particle distribution \citep{Gaburov2011}, is poorly understood.
In general, the dependence of MRI properties on the numerical setup of the 
initial condition should be expected since MRI is extremely sensitive to numerical dissipation. Due to this added complexity
of the initial condition design, although, in principle, MFV-type methods should be less prone to the effect of particle 
discretization noise during slope limiting as well as in the divergence cleaning step, 
we defer their scrutiny in the context of MRI to future work.

We start, in Section \ref{sec:box}, with a discussion of our shearing box implementation and the role of the smoothing kernel function. We also tested the resolution needed for accurate MRI eigenmode growth. In Section \ref{sec:unstrat}, we present unstratified shearing box simulations with and without net vertical flux and in short and tall boxes. Stratified shearing box simulations are described in Section \ref{sec:strat} where we compare different simulation setups and the two methods, MFM and SPH. A discussion and conclusion follow in Section \ref{sec:dis} and \ref{sec:conclusion}.

\section{The Shearing Box Approximation}
\label{sec:box}
The shearing box is a local model of the equations of motion widely used in MRI simulations to achieve high resolution \citep{Goldreich1965, Hawley1995, Latter2017}. One considers a small patch of a disk centered at a radius $R$  and rotating at the angular velocity $\Omega(R)$. In the corotating frame one installs a Cartesian geometry at the box centre, using $x$ and $y$ to represent the radial and the azimuthal directions respectively. In compressible ideal MHD, the governing equations are
\begin{align}
  \frac{\partial \rho}{\partial t}+&\bm{\nabla}\cdot(\rho \bm{v})=0,\\
  \frac{\partial \bm{v}}{\partial t}+\bm{v}\cdot\bm{\nabla}\bm{v}&=-\frac{1}{\rho}\bm{\nabla}(P+\frac{B^{2}}{8\pi})+\frac{(\bm{B}\cdot\bm{\nabla})\bm{B}}{4\pi\rho} \nonumber \\
                                      &+2q\Omega^{2}x\hat{\bm{x}} -\Omega^{2}z \hat{\bm{z}}-2\bm{\Omega}\times\bm{v}, \label{eq:eom} \\
  \frac{\partial \bm{B}}{\partial t}&=\bm{\nabla}\times(\bm{v}\times\bm{B}),  \\
  \frac{\partial \rho u}{\partial t}& +\bm{\nabla}\cdot (\rho u\bm{v})=-P\bm{\nabla}\cdot\bm{v},
\end{align}
where $\hat{\bm{x}}$ and $\hat{\bm{z}}$ are the unit vectors in the $x$ and $z$ directions, and $\rho$, $u$, $P$, $c_s$, $\bm{v}$ represent the density, specific internal energy, gas pressure, sound speed and velocity respectively. The tidal acceleration $2q\Omega^{2}x\hat{\bm{x}}$ in equation \ref{eq:eom} comes from the expansion of the effective potential (gravitational plus centrifugal). The constant $q\equiv -d\ln\Omega/d\ln R$, and for a Keplerian disk $q=1.5$. The vertical component of the star's gravity is represented by $-\Omega^{2}z\hat{\bm{z}}$ which, if included, results in a vertical density stratification with a scale height of $H=c_s/\Omega$, where $c_s$ is the initial sound speed. In simplified models examining motions confined near the disk midplane this term can be dropped. The ratio between the gas pressure and magnetic energy, $\beta\equiv P/(B^{2}/8\pi)$, is a dimensionless measure of the magnetic field strength.

We assume an ideal gas equation of state (EOS),
\begin{equation}
P=\rho u (\gamma -1).
\end{equation}
We choose $\gamma=5/3$ except when we set $\gamma=1.001$ to mimic an isothermal EOS. In particular, we have one stratified simulation with $\gamma=1.001$ to show how such a soft EOS exacerbates long-term numerical dissipation.

\subsection{Shearing Box Boundary Conditions}

 The computation domain is a rectangular prism with sides of length $L_{x}$, $L_{y}$ and $L_{z}$. In unstratified boxes, the domain is periodic in $y$ and $z$, and shear periodic in $x$.
 These boundary conditions can be expressed mathematically for a fluid variable $f$ as
\begin{align}
  f(x,y,z)&=f(x+L_{x},(y-q\Omega L_{x}t)\,\text{mod}\, L_y,z), \\
  f(x,y,z)&=f(x,y+L_{y},z), \\
  f(x,y,z)&=f(x,y,z+L_{z}).
\end{align}
They apply to the thermodynamic variables and the perturbed fluid velocity, where the background is $\bm{v}_0=-q\Omega x\hat{\bm{y}}$.
The implementation of the shearing periodic boundary conditions in Lagrangian codes is relatively easy since we do not need to extrapolate fluid quantities to ghost zones, as in grid-codes. When a fluid element (``particle'') moves across the  radial boundary it reappears at the other radial boundary with a velocity offset added to its azimuthal velocity.

In vertically stratified simulations we apply outflow boundary condition in the $z$ direction by removing any element whose smoothing length is larger than 1.2H. This yields a density floor of about 0.0002 in code units (see below).

\subsection{Equilibrium Tests and the Kernel Function}
\label{sec:pair}
The shearing box admits the following simple equilibrium: $\bm{v}=-\bm{v}_0$, $\rho=$ constant. To test whether the code properly describes this state, in addition to the shearing periodic boundary conditions, we conduct an MFM simulation using this as an initial condition. We employ a cubic box of one disk scale height per side. In the calculation we choose units so that $\Omega=1$, $c_{s}=1$, and $\rho=1.$ At a resolution of $48 \times 48 \times 48$ elements with the Wendland C4 kernel (200 neighbours, $N_{ngb}=200$), the equilibrium can be maintained to machine precision for the duration of the simulation ($\sim 200\Omega^{-1}$) with no sign of breakdown.

\begin{figure}
  \plotone{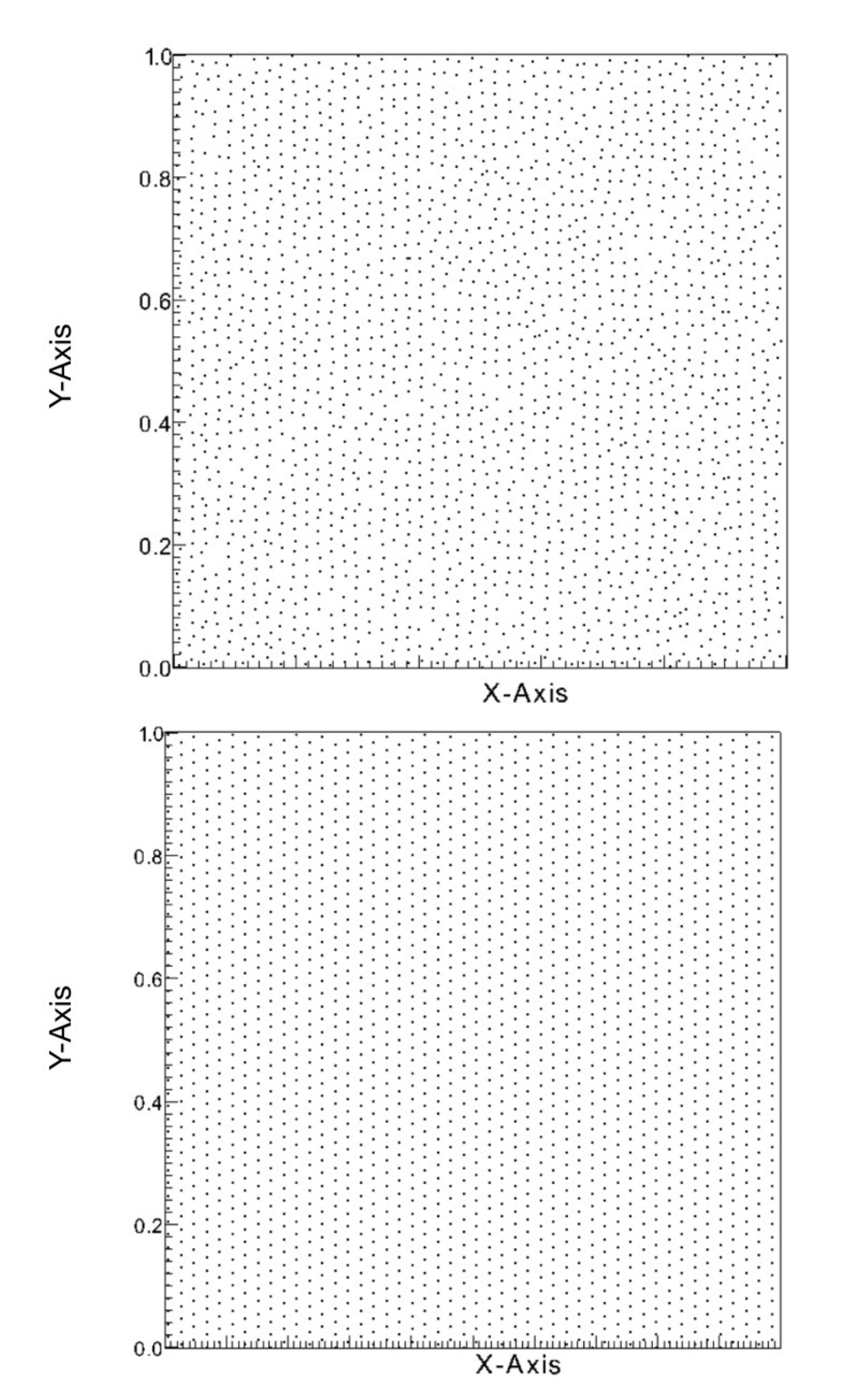}
  \caption{Resolution element (``particle'' or ``mesh-generating point'') locations at $t= 8.4 \Omega^{-1}$ in a \emph{steady state} MFM run with the cubic spline kernel (top) and the Wendland C4 kernel (bottom). Elements form pairs in the simulation using the cubic spline kernel while the Wendland C4 kernel simulation  maintains nice element order (initially cubic lattices are sheared)  \label{fig:pair}}
\end{figure}

We next reran the simulation using the cubic spline kernel (55 neighbours) and found that the radial velocity becomes \emph{non-zero} and the perfect lattice breaks into a glass configuration. 
The velocity errors are a few percent of the sound speed. In this case, elements form pairs as shown in 
figure \ref{fig:pair}. In SPH this pairing (or clumping) instability \citep{Springel2010,Price2012} happens with any kernel whose Fourier transform is negative for some wave vectors, at sufficiently large neighbour number \citep{Dehnen2012}. It would appear then that MFM also suffers a similar instability if ``too many elements'' are included in the kernel radius of compact support (i.e.\ if one does not, as one should, use higher-order kernels with a larger number of neighbors).\footnote{The interpretation of the pairing instability in MFM is slightly different to that in SPH -- the kernel function is used in MFM to define the volume partition between neighboring resolution elements. If one takes a low-order kernel, say, the cubic spline, and forces its radius of compact support to enclose too many elements (such that the mean inter-element separation is much smaller than the kernel function width), the effective faces between elements essentially ``overlap'' into a single face (which becomes ill-defined).} The Wendland C4 kernel does not suffer from these issues at this ``enclosed neighbor number'' \citep[see e.g.][]{Dehnen2012}. It also helps to keep elements well ordered, which is crucial for accurate gradient estimation in any unstructured method \citep{Rosswog2015}. MRI turbulence is generally subsonic (except in the disk corona of stratified box), so we always use the Wendland C4 kernel to minimize numerical noise/dissipation (except when noted).

\subsection{Resolution}
\label{sec:resolution}

Ideally we would like to estimate the shortest lengthscale adequately resolved in a simulations $h$, the number of resolution elements required to furnish that level of accuracy, and ultimately the numerical resources required by a given code 
to achieve this. 

Unfortunately, it is not straightforward to determine the shortest resolvable lengthscale, even for grid codes (which possess an unambiguous grid spacing $\delta z$). Particle codes exhibit several lengthscales that might be used
as references in this context. Perhaps the most obvious choice is the
mean particle separation $\overline{\delta}$. Of course, this is rather crude, because particle codes are intrinsically adaptive, with denser regions better resolved than less dense regions.
But even if we put this concern to one side, we must
emphasise that resolution is not so
much determined by the size of $\overline{\delta}$ as it is
by the kernel volume, since this is where the interpolation occurs (which is instrumental in defining a volume element in MFM). We may then be tempted to use, instead of $\overline{\delta}$, the kernel support radius $H_1$, because the particles within a kernel form a `resolution unit' together.
But more accurate estimates exist, as we now discuss.

In SPH, an alternative definition for the resolution scale is $h=2\sigma$ where $\sigma$ is the standard deviation of the weighting kernel function $W(\bm{x},H_1)$, as defined in \citet{Dehnen2012},
\begin{equation}                                                                                           \sigma^{2}=\frac{1}{3}\int d\bm{x}x^{2}W(\bm{x},H_1).                                                   \end{equation}
We hence will denote this lengthscale by $h_\text{SPH}$.

In MFM, for a well-chosen kernel (i.e. $H_1$ is within a factor of a few rms inter-particle separation, so that faces are well-defined), an alternative estimate of $h$ is the face-area weighted inter-neighbor separation (where the face areas are themselves determined by the volume partition from the kernel function; \citealt{Hopkins2015a}). Note that, consequently, h so defined is the scale upon which the Riemann problem is solved, diferent to both $H_1$ and $\overline{\delta}$. For the Wendland C4 kernel parameters adopted here, this yields $h \approx 0.45 H_1$, so we will use this value throughout. We denote this by
$h_\text{MFM}$. 

\subsection{MRI Quality Factor}
\label{sec:quality}

In grid-code MRI simulations, the number of cells per fastest growing mode's wavelength is an often used resolution metric \citep{Hawley2011, Parkin2013}. It can potentially tell us where and when in a simulation the MRI is adequately or inadeqately resolved. 

We may define a quality parameter \citep{Noble2010, Hawley2011} via
\begin{equation}\label{e:qfactor}
  Q_z=\frac{\lambda_{MRI}}{h}=\frac{2\pi V_{az}}{\Omega h},
\end{equation}
where $V_{az}$ is the $z$ component of the Alfven velocity and $h$ is the (vertical) resolution length. $\lambda_{MRI}$ is close to but not exactly the fastest growing linear mode's wavelength, $\lambda_{fastest}=\sqrt{16/15}\lambda_{MRI}$, in the presence of a net vertical flux. Perturbations with wavelengths smaller than $\lambda_{MRI}/\sqrt{3
}$ are stable for the same configuration.  $Q_z$ can be measured and averaged over the disk body during the saturated state. It can also be defined with respect to other coordinate directions,
most often in the $y$ direction. Regions where the plasma beta is high are of lower quality factor, and in these regions there may be insufficient resolution. 

We must state from the outset that a quality factor, so defined, is a rather crude measure of how well the turbulence is resolved. First, it is based
on the linear theory of the net-flux MRI set-up and hence may not be generally applicable during the ensuing nonlinear saturation; certainly its relevance for zero-net flux simulations is debatable. Second, it only describes whether the input scale of the turbulence is resolved and has nothing to say about the ensuing turbulent cascade on smaller scales. If $Q_z \gtrsim 1$ then there is no inertial range to consider, and, in addition, the input and dissipative scales are directly adjacent: strictly, there is no real turbulence, but rather a monoscale chaotic flow. Nonetheless, vertically stratified shearing box simulations indicate  that $Q_{z}  > 10, Q_y  > 20$ ensures the convergence with resolution of some large-scale average flow quantities \citep{Hawley2011}. 

In our small number of grid-based simulations we substitute
$h=\delta z$, the grid spacing. In our particle simulations we 
use both $\overline{\delta}$ and $h_\text{SPH}$ or $h_\text{MFM}$ for comparison. We denote by $Q_\text{sepr}$ the quality factor when $\overline{\delta}$ is used for h, and denote by $Q^*$ the quality factor when either $h_\text{SPH}$ or $h_\text{MFM}$ is used. 

For MFM the two quality factors are related by
\begin{equation}                                                                 \frac{Q^*}{Q_\text{sepr}}=(\frac{4\pi}{3N_{ngb}})^{\frac{1}{3}}\kappa,                                              \end{equation}
where $\kappa \equiv H_1/h_\text{MFM}$ and is related to the type of kernel used in the simulation. 
The cubic spline kernel can achieve good density estimation with a small number of neighbours $N_{ngb}$ (around 42), where we define
\begin{align}
  N_{ngb}&=\frac{4\pi}{3}H_{1}^{3}n(\bm{x}_{i})=\frac{4\pi}{3}H_{1}^{3}(\rho_{i}/m_{i}),\\
         n(\bm{x}_{i})&=(\overline{\delta})^{-3},
\end{align}
where $m_i$ is the mass of the $i$'th particle.
The cubic spline yields $Q^*$ and $Q_\text{sepr}$ that are relatively close (the latter only a factor 0.9 of the other). But the Wendland C4 kernel  ($N_{ngb}=200$) gives $Q^*=0.6Q_\text{sepr}$. 



\subsection{Channel flow growth}
\label{sec:channel}

The linear MRI eigenmodes in a net vertical flux are called channel flows. Being nonlinear solutions,
these eigenmodes will grow to large and nonlinear amplitudes before being destroyed, in the first instance, by parasitic instabilities \citep{Goodman1994}. A robust turbulence then ensues, which sometimes exhibits the recurrent generation and destruction of the channels. 
In this section we measure the growth rates of the simulated channel flow in SPH and MFMand compare it with the theoretical value. 
In particular, we calculate the error as a function of resolution (in fact by, $Q$), to see if certain scalings
can be discerned. For comparison we also obtain error estimates
for the finite volume Godunov code, ATHENA \citep{Stone2010}, with second order reconstruction and either the Roe and HLLD solvers. Our primary aim here is to compare the performance of the codes on this linear problem not with respect to computational effort directly, which we address in detail in Section \ref{sec:hours}, but with respect to $h$. Is a plausible estimate of the minimum resolvable lengthscale a useful diagnostic for measuring the accuracy of MRI growth, or are other things going on? If the importance of $h$ is overwhelming we may in fact expect all three codes to perform similarly on this test. 

\begin{figure}[ht!]
  \epsscale{1.2}
  \plotone{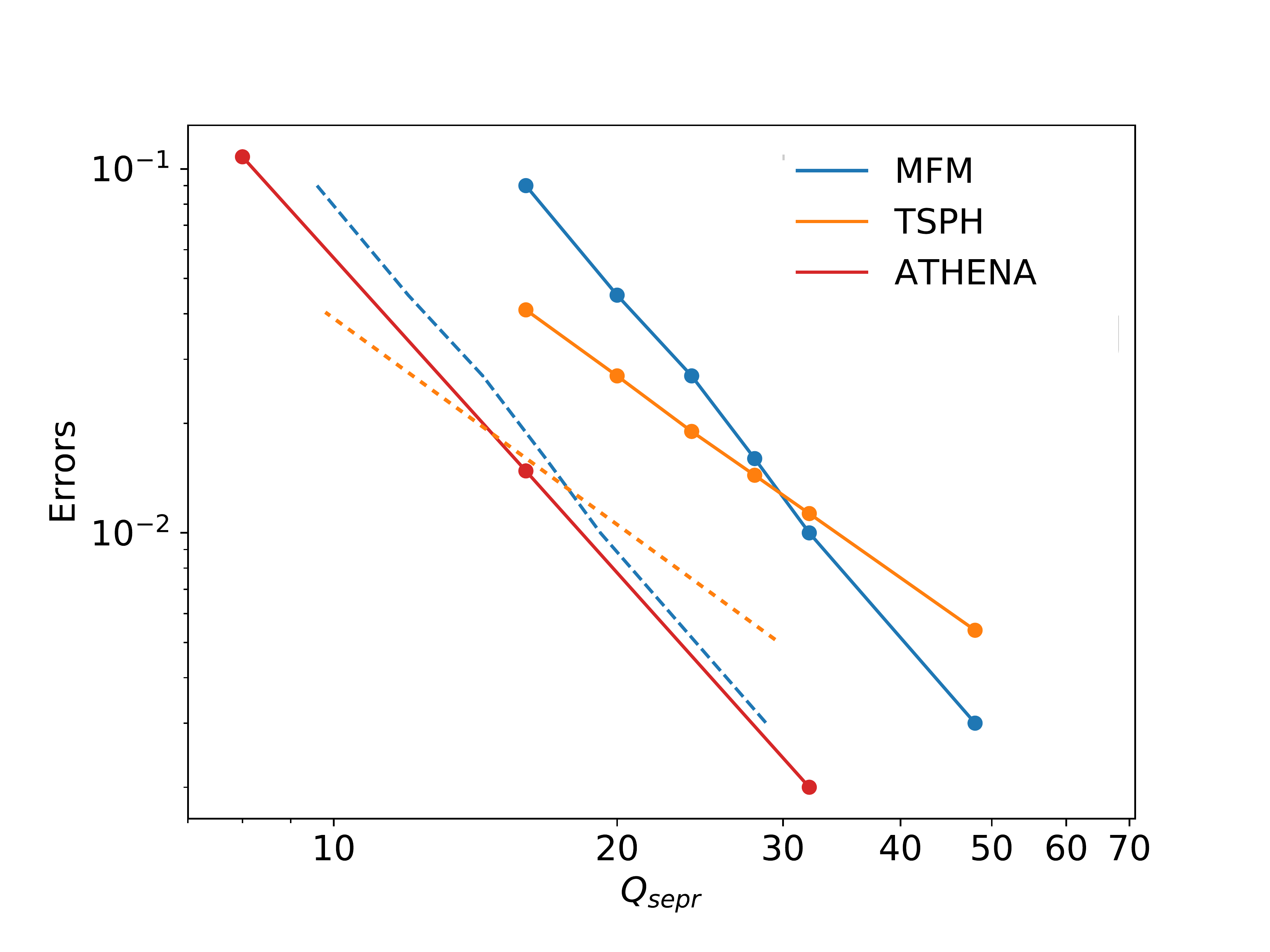}
  \caption{The growth rate errors in different MHD schemes versus the quality factor $Q$ (number of resolution lengths $h$ per MRI wavelength) for various codes and different definitions of $h$. The solid red line indicates 
  the performance of the ATHENA code for three different
  resolutions; here $h=\delta z$. The solid blue line represents MFM and the solid orange curve TSPH, both using the mean particle separation $\overline{\delta}$ for $h$, i.e. $Q=Q_\text{sep}$ and equals
  the mean number of particles per MRI wavelength. The dashed lines represent the performance of MFM and TSPH when $Q=Q^*$ (see Section \ref{sec:quality}). 
   \label{fig:gr}}
\end{figure}

We initialize a box of size $H \times H \times H$, threaded by uniform vertical background fields of magnitude $B_{0}$.  We set $\beta=84$, so that the fastest growing channel mode just fits into the box, and $\gamma= 1.001$ so the gas is effectively isothermal. The initial amplitude of the channel mode is $0.001B_{0}$. The theoretical growth rate of the fastest channel mode is $0.75\Omega^{-1}$. The simulations are run for $8\Omega^{-1}$ so that the channel mode is smaller than $0.4B_{0}$ at the end of the simulation. We calculate the growth rate  using the magnitude of the magnetic field at two consecutive snapshots taken every $0.5\Omega^{-1}$. The growth rate relative error is defined as
\begin{equation}
  \text{max}\{(s_{i}-0.75)/0.75\}
\end{equation}
where $s_{i}$ is the $i$th measure of the growth rate. 

In figure \ref{fig:gr}, we plot the growth rate error as a function of $Q_\text{sepr}$, i.e. the (mean) number of resolution elements per $\lambda_{MRI}$, be they grid cells (in ATHENA) or the average particle spacing (in SPH and MFM). In MFM the Wendland C4 kernel is employed in order to avoid pair instability. As is clear, and to be expected, the errors decline with increasing resolution. They are less than 1\% when $Q_\text{sepr}>32$ in MFM. TSPH captures the MRI better than MFM in the low resolution simulations but it converges slower. It is known that SPH has zeroth order errors that only vanish when both $Q_{sepr}$ and $N_{ngb}$ approach infinity ($N_{ngb}$ is fixed here) \citep{Read2010,Zhu2015}. 

ATHENA performs better than both SPH and MFM, if we are to directly compare number of grid points against number of particles. But, as explained in the previous two subsections, even if such a comparison may reflect relative numerical effort it doesn't do justice to the particle codes, for whom the mean particle spacing is not equal to the shortest resolvable lengthscales. 
In figure \ref{fig:gr} we also plot the relative growth rate error for both SPH and MFM versus $Q^*$, the number of resolution lengths $h$ per MRI wavelength (less than
the number of particles). These dashed curves now sit on top
or adjacent to each other. If we are to consider resolution this way, actually the codes perform comparably; ATHENA and MFM (which share the same Riemann solver) agree exceptionally well. In contrast, SPH possesses a different and adverse scaling in comparison, indicative of its sensitivity not only to the number of particles but to the number of near neighbours. This poorer performance we will see again later in nonlinear tests of MRI saturation.

Finally, we say a few words dealing with the cubic spline kernel in MFM, which for smaller $N_{ngb}$ yields a larger effective quality parameter $Q^*$ (see section \ref{sec:resolution}). Indeed, it outperforms the Wendland C4 kernel when the resolution is low. However, when $Q_{sepr}>20$ the growth rate errors increase due to the pairing instability (which itself is resolution dependent, see \citet{Dehnen2012}). This numerical noise (see section \ref{sec:pair}) can dominate over the weak channel mode in the early stage. The channel modes does eventually outcompete this noise but the errors in gradient estimation lead to extra dissipation (see section \ref{sec:sph}) which is hard to quantify.

\section{Unstratified Shearing Box Simulations}
\label{sec:unstrat}

\begin{deluxetable*}{c|c|c|c|c|c|c|c}[ht!]
  \tablenum{1}
  \tablecaption{Simulations, results and comments \label{t:simulations}}
  \tabletypesize{\scriptsize}
  \tablewidth{\textwidth}
  \tablehead{
    \colhead{Simulations} & \colhead{Initial fields} & \colhead{Boxsize} & \colhead{Resolution} & \colhead{MHD-methods} & \colhead{EOS} & \colhead{Sections/Figures} & \colhead{Ref}}
  \startdata
  \multirow{4}{*}{Unstratified}& \multirow{2}{*}{NZ} & $H\times 6.28H \times H$ & $64\times 360 \times 64$ & MFM & \multirow{2}{*}{adiabatic} &\multirow{2}{*}{Sec \ref{sec:nz}/Fig \ref{fig:nz}}  & \multirow{2}{*}{1} \\
                               &                         & $H\times 4H \times H$    & 48 elements per H                     & MFM/PSPH &                      &           &                    \\
                               & \multirow{2}{*}{ZNZ}    & $H\times \pi H \times H$ & $64\times 200 \times 64$ & MFM        & isothermal& Sec \ref{sec:znz}/Fig \ref{fig:znz} & 2                 \\
                               &                         & $H\times 4H \times 4H$   & 48/64 elements per H & MFM           & adiabatic(c) & Sec \ref{sec:znzt}/Fig \ref{fig:tall}  & 3        \\
\hline
  \multirow{3}{*}{Stratified} & \multirow{3}{*}{$B_y$}  & \multirow{3}{*}{$\sqrt{2}H\times4\sqrt{2} \times 24 H$} & \multirow{2}{*}{1.5M elements}&TSPH/PSPH  & adiabatic & Sec \ref{sec:sph}/Fig \ref{fig:sphmag}  & \multirow{3}{*}{4} \\
                              &                         &                                                    &                                & MFM & ad/iso & Sec \ref{sec:mfm}/Fig \ref{fig:mfmmag}   \\
                              &                         &                                                    & 3M elements                   & MFM & adiabatic(c)    & Sec \ref{sec:3M}/Fig \ref{fig:3M}  \\
\enddata

\tablecomments{
 The following abbreviations have been used: \textbf{NZ} - Net vertical flux; 
\textbf{ZNZ} - Zero net vertical flux; \textbf{MFM} - meshless finite mass method with constrained gradient divergence cleaning; \textbf{TSPH} - Density-energy (traditional) formulation of SPH; \textbf{PSPH} - Pressure-energy formulation of SPH. Here we always uses the Wendland C4 kernel except one experiment run with the quartic spline kernel in figure \ref{fig:mfmmag}.
In stratified shearing box simulations, elements with smoothing length larger than 1.2H are clipped resulting in a density floor $\sim 0.0002$. Both SPH MHDs employ the Cullen \& Dehnen artificial viscosity switch \citep{Cullen2010}, the hyperbolic divergence cleaning of \citet{Tricco2012}, and the artificial resistivity of \citet{Tricco2013}. Furthermore,
\textbf{adiabatic} runs use $\gamma=5/3$, \textbf{isothermal} runs $\gamma=1.001$, and \textbf{adiabatic(c)} runs apply an \emph {ad hoc} cooling (see equation \ref{eq:cool}).
We expect very fast turbulence decay due to numerical dissipation using isothermal EOS (see figure \ref{fig:mfmmag} \& section \ref{sec:mfm}). We always try to use an adiabatic EOS to minimise long-term numerical dissipation except when we want to enable direct comparison with previous studies.\\
Ref. 1. \citet{Hawley1996}, 2. \citet{Fromang2007a}, 3. \citet{Shi2015}, 4. \citet{Davis2010}}
\end{deluxetable*}

 We summarize all the simulations we undertook in both unstratified and stratified boxes in table \ref{t:simulations}, which includes key parameters, physical and numerical configurations, comments, and references. Further details can be found in the referenced subsections.
 
\subsection{Net-Vertical-Flux Simulations}
\label{sec:nz}

We first ran an unstratified shearing box simulation with net vertical flux similar to the fiducial model of \citet{Hawley1995}. This is the simplest 3D MRI setup.
For such configuration we are able to reproduce the main features of previous grid-based simulations using high resolution MFM simulations. The box is of size $H\times 6.28 H \times H$ and threaded by vertical fields with $\beta=400$. We used a resolution of $64\times 360 \times 64$ elements which corresponds to 28 elements per $\lambda_{MRI}$, and hence roughly 17 $h_\text{MFM}$ per $\lambda_{MRI}$ (or roughly 9 $H_1$
per $\lambda_{MRI}$). In comparison,  \citet{Hawley1995} 
employed about 16 grid points per MRI wavelength. 
The box size also affects the simulated turbulence: smaller box tends to have stronger outbursts in the turbulent state \citep{Bodo2008, Lesaffre2009}.  We set two other runs in a box of $H\times 4H\times H$ with $\beta=330$ ($H=2\lambda_{MRI}$), and with either PSPH and MFM using 48 elements per \emph{H}.
We added random velocity perturbations (5\% of the sound speed) to the shear flow at initialisation. The simulations are run for 11 orbits with $\gamma=5/3$. 

\begin{figure*}[ht!]
  \epsscale{1.2}
  \plotone{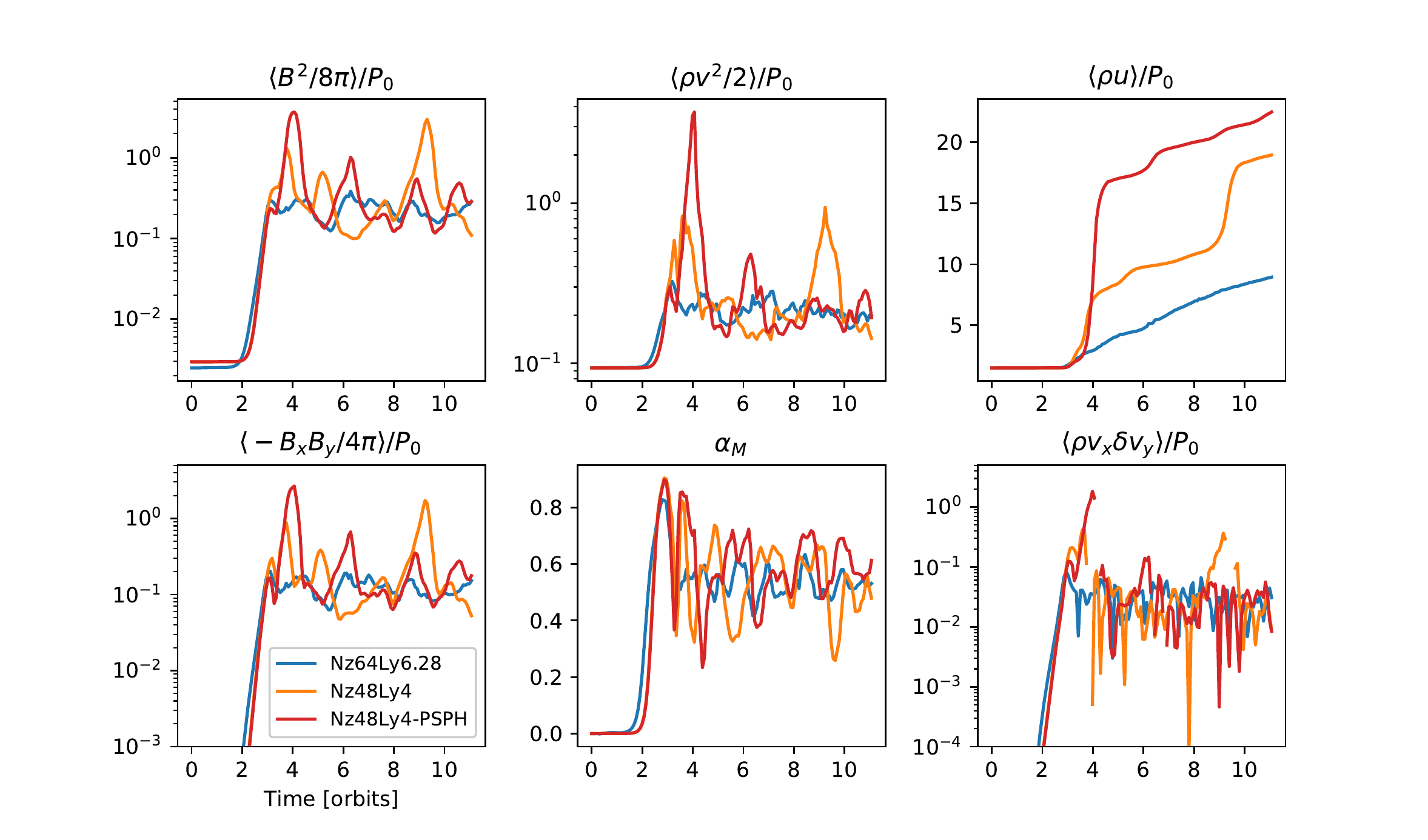}
\caption{From upper left to bottom right corner, the time evolution of the averaged magnetic energy, kinetic energy, thermal energy, Maxwell stress, $\alpha_{M}$ and Reynold stress in the unstratified vertical flux simulations are shown (see text for the
explanation of how average quantities are computed).
Time is given in orbits. $P_{0}$ is the initial pressure. The MFM simulation with $L_y=6.28 H$ (red curves) gives results close to those of \citet{Hawley1996}. The PSPH simulation with $L_y=4H$ (black curves) has larger internal energy than the two MFM simulations. We note that the increasing internal energy leads to a larger plasma $\beta$ and smaller outbursts.\label{fig:nz}}
\end{figure*}

\begin{figure*}[ht!]
  \epsscale{1.2}
  \plotone{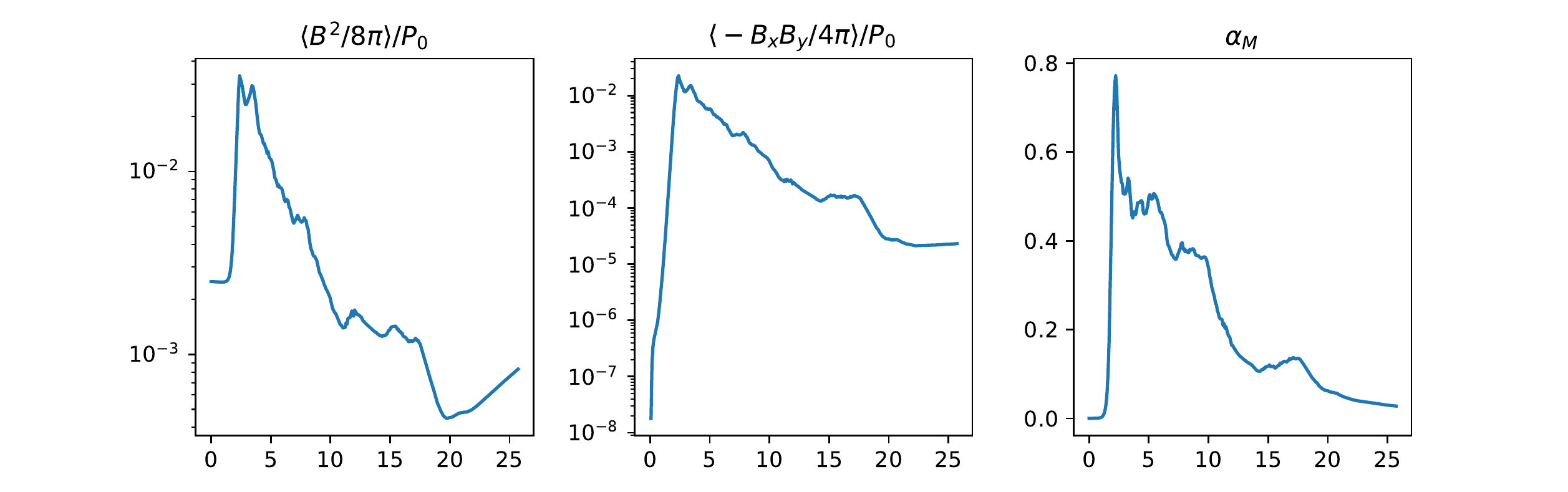}
\caption{Time evolution of averaged magnetic energy, Maxwell stress and $\alpha_{M}$ in the zero net flux MFM simulation. Time is given in orbits. The magnetic field decays quickly and the Maxwell stress becomes nearly zero after about 20 orbits (see text) .\label{fig:znz}}
\end{figure*}

 To characterize the saturated turbulence we plot several density weighted-averaged quantities in figure \ref{fig:nz}. 
 We take the arithmetic average of fluid variables at all the MFM  fluid elements so the average is naturally density-weighted because of the adaptive nature of GIZMO. In unstratified turbulence, the density fluctuations are small so the density-weighted average should be close to the volume average in previous studies.  This should be the case also for stratified turbulence, another situation in which we will apply this
 averaging method (see next section \ref{sec:strat}), because the stress
 is  almost independent of the density when $\vert z \vert <2\sqrt{2}H$ \citep{Simon2011}. In the $L_y=6.28H$ simulation (red curves), both the magnetic energy and stresses are in agreement with the results of \citet{Hawley1995}, which
 were obtained with an Eulerian code.
 The ratio of the Maxwell stress to the magnetic pressure,
\begin{equation}
  \alpha_{M}=\frac{\langle -2B_{x}B_{y}\rangle}{\langle B^{2}\rangle},
\end{equation}
is about $0.5$, similar to the aforementioned
previous adiabatic calculations. The agreement between the older ZEUS runs and MFM may have been expected because the resolution
is similar in the two (about 16 resolution lengths, either $\delta z$ or $h_\text{MFM}$, per MRI wavelength). It should be noted, however,
that the ZEUS code is more diffusive than ATHENA - our point of comparison in the linear growth tests.

The simulations in smaller boxes, $L_y=4H$, show stronger 
bursts in stresses and magnetic energy because fewer active (non-axisymmetric) modes can fit in the box 
leading to an artificial truncation of the  participating
modes in the nonlinear dynamics. As a consequence, the system is nearer criticality and
single channel modes can intermittently dominate.

The internal energy increases due to the turbulent dissipation. This is most significant in the two $L_y=4H$ simulations as they exhibit the strongest bursts from the channel flows: since these flows achieve large amplitudes, when they break down a great deal of energy is dissipated into heat. The PSPH simulation, in particular, is some four times `hotter' than the large box MFM simulation. The PSPH run undergoes also a much greater
increase of internal energy compared to the equivalent MFM run (see Figure \ref{fig:nz}, upper right panel) due to  stronger channel activity near the beginning of the run (signaled by the very large initial spike in the various diagnostics shown in Figure \ref{fig:nz}). The dominance of channels in the PSPH run early on suggests that the system is closer to marginal stability than in MFM; this is probably due to the additional numerical diffusivity in PSPH. Note that the plasma $\beta$ increases substantially as the gas is heated up, and the boxes will ultimately approach the incompressible zero-net-flux regime. This explains why, in general, the bursts become less powerful as the simulations continue. Finally,
for the net-flux unstratified setup we only ran PSPH, and not TSPH;  since there are no 
steep density gradients, we expect no significant difference due to the actual formulation of the SPH hydro force.

\subsection{Zero-Net-Vertical-Flux Simulations in a `Standard' Box}
\label{sec:znz}

 We run a standard zero-net-flux unstratified box simulation with MFM with no explicit physical dissipation terms \citep{Stone1996,Fromang2007a}. We initialize a box of size $H \times \pi H \times H$ with $64\times 200 \times 64$ elements, threaded by magnetic fields,
\begin{equation}
  \bm{B}=B_{0}\hat{\bm{z}}\sin(2\pi x/H).
\end{equation}
The field strength $B_{0}$ is chosen so that the volume averaged $\beta$ equals 400. We use an isothermal EOS ($\gamma=1.001$) to align with \citet{Fromang2007a}. MRI turbulence is sensitive to the nature of both physical and numerical dissipation: without physical viscosity and resistivity, \citet{Fromang2007a} found that zero-net-flux MRI turbulence was driven to smaller scales as resolution increased with no sign of convergence.
\citet{Fromang2007b} showed that the saturated state depended on the magnetic Prandtl number when a source of diffusivity, physical or numerical, is present; if this was too low, turbulence would die after some period of time. With these problematic results in mind, we will now assess how well a zero-net-flux MRI setup can be modelled by 
a meshless code.

\begin{figure*}[ht!]
\epsscale{1.2}
\plotone{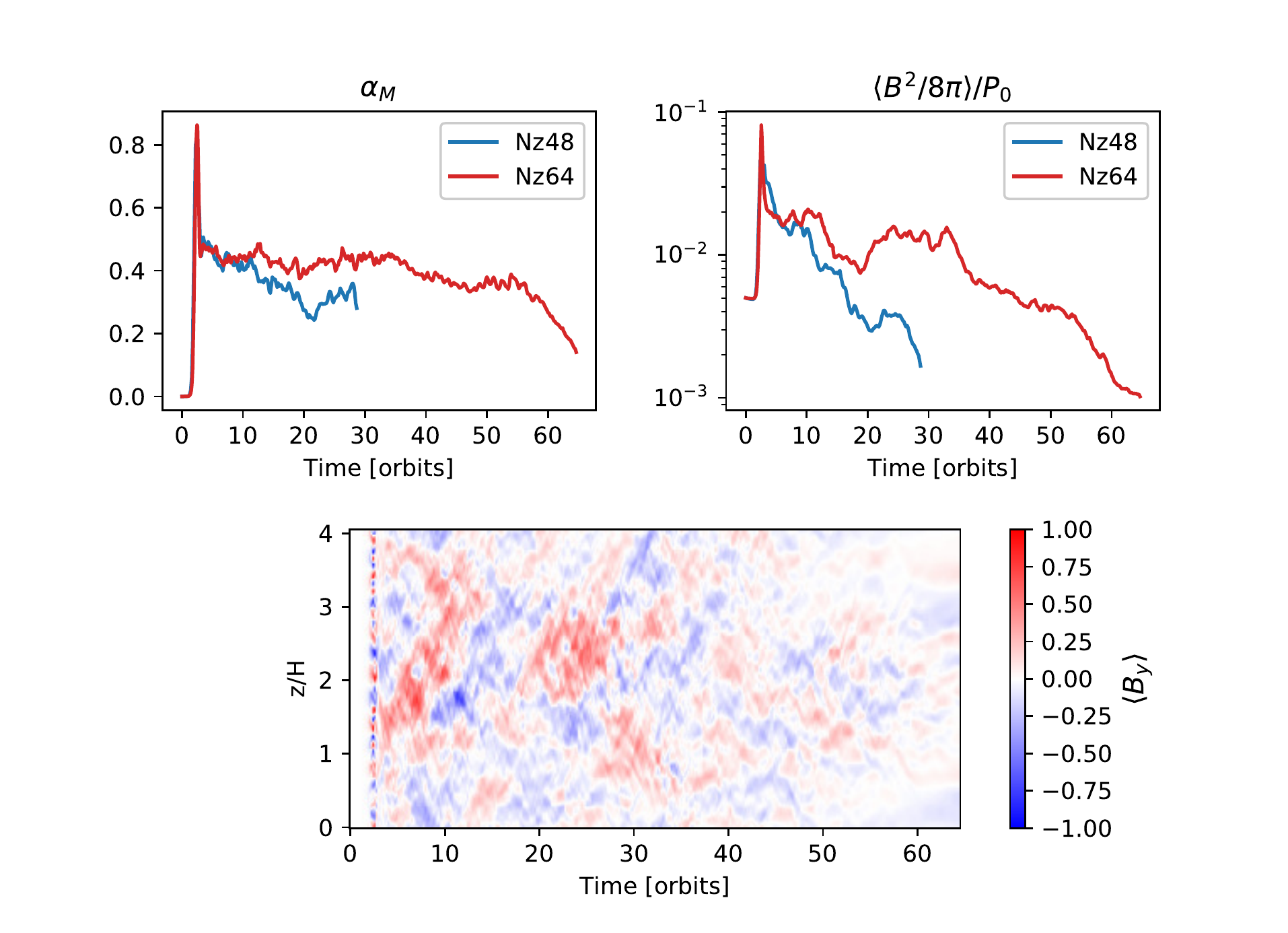}
\caption{The time evolution of $\alpha_M$ and averaged magnetic energy in the zero net vertical flux tall box MFM simulation are shown in the upper panels. In the lower
panel the temporal evolution of the averaged horizontal magnetic field of the Nz64 simulation is presented.
The simulation (Nz64) with 64 elements per scale height shows a sign of convergence comparing to the fast decay of magnetic fields in the short box zero net flux simulation in figure \ref{fig:znz}. The pattern of the averaged azimuthal field is also similar to that of \citet{Shi2015}. However, the magnetic fields eventually decays.\label{fig:tall}}
\end{figure*}

We plot the averaged magnetic energy, Maxwell stress, and $\alpha_{M}$ in figure \ref{fig:znz}. In contrast to MRI runs with grid codes, after an initial burst the magnetic fields and magnetic stress rapidly decay. There is no sustained turbulence, as in \citep{Fromang2007a} nor is there some period of MRI turbulence before decay, as in \citet{Fromang2007b}. It is true that MFM smooths fluid variables within a kernel, so the resolution (with respect to $h_\text{MFM}$) is roughly half that of the standard simulation with $64\times 200 \times 64$ grid cells in \citet{Fromang2007b} (see section \ref{sec:resolution}). A simulation with $>$128 particles per scale height is prohibitively expensive with MFM (see section \ref{sec:hours}). But even an `ideal MHD' run undertaken with low resolution in a  grid code can sustain MRI turbulence \citep{Stone1996}. Thus our result is disappointing.

  One way to interpret it is to consider the relative sizes of the numerical resistivity and viscosity.
  At low resolutions, MFM should have a moderate numerical viscosity, $\nu$ (see appendix \ref{sec:nu}) and relatively large numerical resistivity, $\eta$ (see appendix \ref{sec:eta}), as a consequence the effective numerical Prandtl number $P_{m}=\nu / \eta$ must be small (smaller than 1). It is then perhaps not surprising that the turbulence decays \citep{Fromang2007b}. However, the fact that it decays so abruptly might point to a simpler reason: the numerical resistivity is just very large and prohibits
  turbulence of any kind past the initial spike. Indeed, our MFM run resembles in some respects the \citet{Fleming2000} run with a magnetic Reynolds number of 13000, which after an initial burst abruptly dies off.

\subsubsection{Zero-Net-Vertical-Flux Simulations in a Tall Box}
\label{sec:znzt}

It has been shown that when the numerical domain is reshaped, so it exhibits a large vertical aspect ratio ($L_z/L_x \geq 2.5$), a new more vigorous and cyclical MRI dynamo emerges \citep{Shi2015} (see also \citet{Lesur2008}). Importantly, its saturated stress is independent of resolution. To test the effect of a taller box, we redo the simulations in Section \ref{sec:znz} in a box of size $H\times 4H \times 4H$. We present two simulations with 48 and 64 elements per scale height; their details are shown in figure \ref{fig:tall}. We set $\gamma=5/3$ and add a cooling term in order to keep the
internal energy roughly constant (see equation \ref{eq:cool}). The numerical dissipation is EOS-related and we expect very fast decay with the isothermal EOS (see section \ref{sec:mfm}). This does mean, however, that we can't make direct quantitative comparison with \citet{Shi2015}.

Initially our simulations exhibit turbulence as shown in Figure \ref{fig:tall}. Moreover they reproduce the averaged toroidal field patterns produced by \citet{Shi2015}. However, while the turbulence is sustained for much longer than in the standard box, 
after some 30-40 orbits the activity ultimately dies out. During the turbulent phase,
$\alpha_M \sim 0.44$ in the higher resolution tall box simulation, but the averaged magnetic energy $\langle B^{2}/8\pi \rangle / P_0 \sim 0.01$ is much lower than the values ($>0.1$) obtained with the ATHENA code \citep{Shi2015}. As a result, the stress ($\sim 0.006 P_0$) is also much smaller.

In tall box simulations, when the magnetic Prandtl number, $P_m \geq 4$ the saturated stress is independent of $P_m$ while the turbulence vanishes for $P_m=1$ with even 128 cells per scale height \citep{Shi2015}. While $P_m$ effects might be at play in our simulations it should be noted that our 64 particle simulation is quite low resolution (elements are further smoothed within a kernel). Worse resolution leads to more rapid decay. It is likely that both meshless simulations possess too great a numerical resistivity to support a sustained MRI dynamo.

To summarize, the MFM simulations -- at least with the current implementation of the
method -- do not allow us to sustain the MRI in unstratified zero-net flux simulations, 
either in short or tall boxes. This is probably the result of either too large a numerical resistivity, or, more generally, too low an effective $P_m$, at least at the resolutions we were able to access (see Section \ref{sec:hours}). It will be particularly interesting to explore both (a) higher-resolution MFM simulations (where the resistivity should be lower and $P_{m}$ larger), and (b) simulations using other, closely-related schemes which are not completely ``fixed mass'' schemes but closer to moving mesh schemes (e.g. the MFV scheme or arbitrarily shearing-mesh schemes with the constrained-gradient divergence cleaning). 
We did not run zero-net-flux simulations with SPH, because of
evidence suggesting that it has
an intrinsically larger numerical viscosity, and thus would yield higher Prandtl numbers, quite irrespective of the specific implementation of artificial viscosity \citep{Bauer2012,Deng2017a,Deng2017b}, and produces substantially larger magnetic field divergence with available divergence cleaning methods (see Section \ref{sec:divb}). 

\begin{figure*}
  \plotone{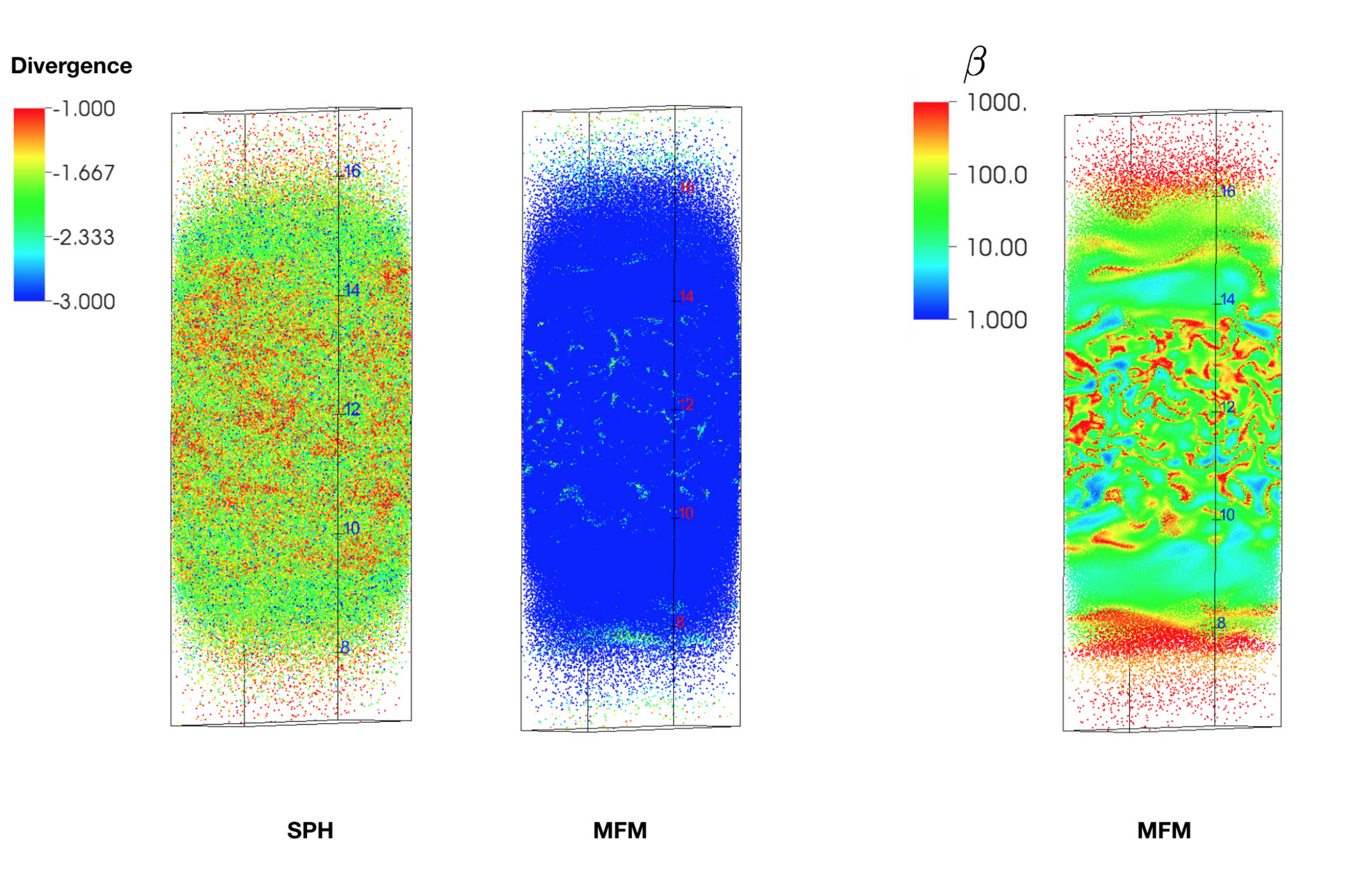}
  \caption{Typical $Log_{10}(divB)$ value in TSPH (PSPH shows similar results) and MFM. MFM with CG divergence cleaning maintains 2 orders of magnitudes smaller $divB$ than the hyperbolic divergence cleaning 
  scheme used in TSPH. In MFM, most elements have $divB<0.01$. Large $divB$ only occurs in the weak field ($\beta>1000$) regions.\label{fig:divb}}
\end{figure*}

\section{Stratified Shearing Box Simulations}
\label{sec:strat}

In this section we undertake simulations in the vertically stratified shearing box, in which the 
(leading order) vertical acceleration from the central star's gravity is incorporated. We initialise the simulations with a Gaussian density profile with uniform temperature
\begin{equation}
  \rho(z)=\rho_{0}\text{exp}(-\frac{z^{2}}{2H^{2}}),
\end{equation}
where $H=c_{s}/\Omega$ is a factor different from $\sqrt{2}c_{s}/\Omega$ in some previous work \citep{Davis2010, Simon2011}. Note that $c_s$ is the initial sound speed; in adiabatic runs the sound speed (and hence scale height) will change. We adopt units so that $H=1,\Omega=1,c_{s}^{2}/ \gamma=1$ and use the adiabatic EOS with $\gamma=5/3$. The density profile is sampled using the Monte Carlo method and then relaxed to a glassy configuration. The density errors in the disk body ($-3H<z<3H$) are below the 1\% level. We initialize azimuthal magnetic fields with $\beta=25$ in a box of size $\sqrt{2} H\times 4\sqrt{2} H\times 24H$ (the box is extremely tall but no element has $\vline z \vline >6$ in our simulations, see figure \ref{fig:MF}). Outflow boundary condition are applied but in fact there is no significant outflow and few elements are clipped. Random velocity perturbations $\sim 0.01c_s$ are added to seed the instability.

In our fiducial model we use 1.5M elements in total, leading to $\overline{\delta}\sim 0.03$ at the disk midplane. However, due to the adaptive feature of our method, the resolution is lower the further away from the midplane. This helps in order to 
save some computational resource because high resolution is not needed
in the MRI-stable disk corona with strong fields \citep[see figure \ref{fig:MF} and also][]{Miller2000}. Yet, the nearly 
zero-flux MRI turbulence in the disk body still requires high resolution and is computationally demanding.

\begin{figure*}
  \epsscale{1.2}
  \plotone{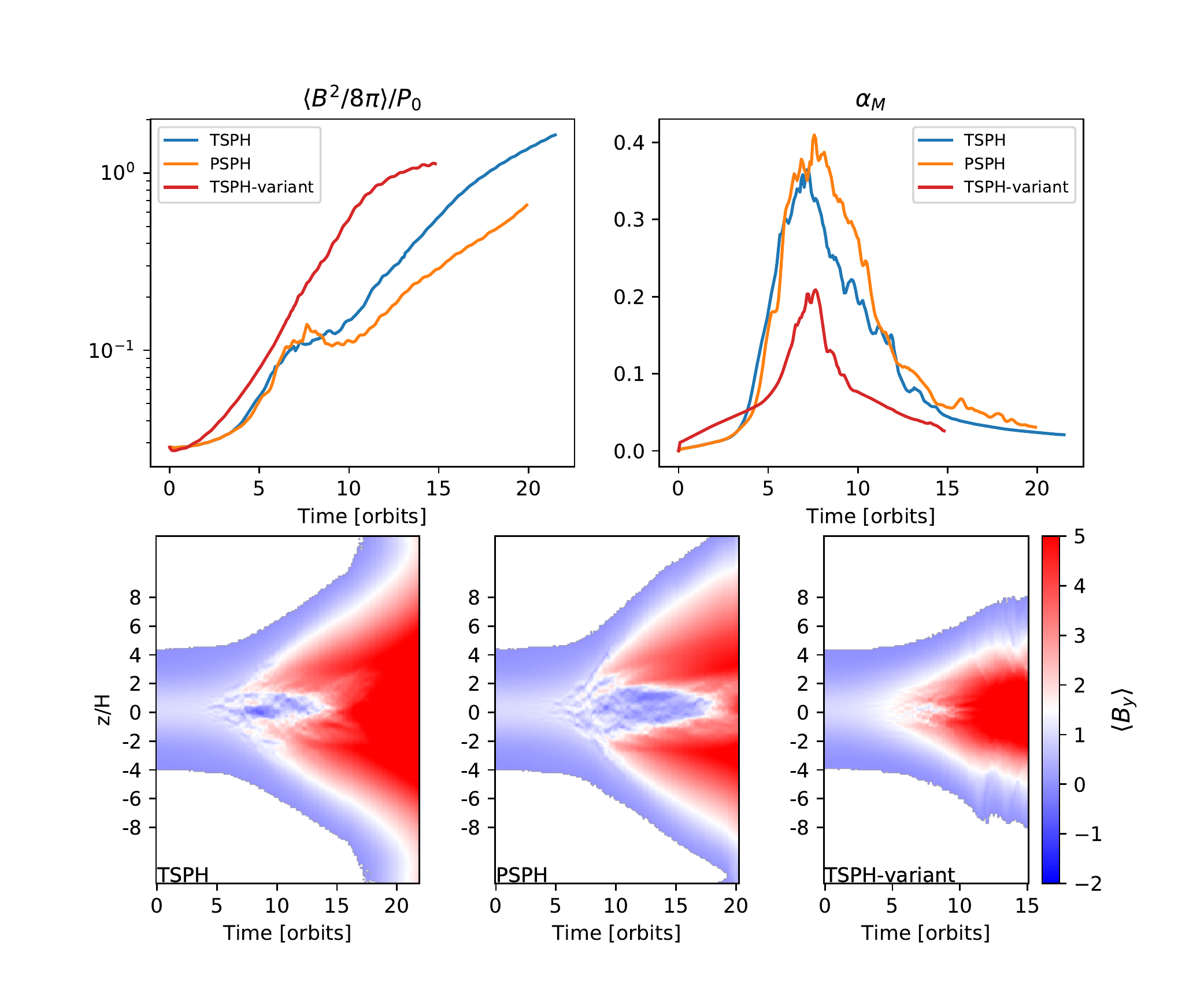}
  \caption{The evolution of the magnetic fields in the SPH MHD stratified shearing box simulations.  The upper panels are the time evolution of the magnetic energy and $\alpha_M$ as noted. The lower panels show the time evolution of the horizontally averaged azimuthal magnetic fields. Strong toroidal fields grow through shear amplification of radial fields. Secondary instability cannot develop efficiently in the low resolution disk corona and the strong toroidal fields spread gradually to the disk midplane. The stratified box is eventually filled with strong toroidal fields ($\beta \sim 1$, stable to MRI) and the disk expands vertically. The PSPH and TSPH simulation are almost identical since their artificial viscosity damps subsonic turbulence similarly \citep{Bauer2012, Hopkins2015a}. We add a more dissipative TSPH simulation, with the quartic spline kernel and isothermal EOS (see figure \ref{fig:mfmmag}). The TSPH-variant simulation doesn't dissipate the strong toroidal fields but grows the fields even quicker due to larger numerical noise. \label{fig:sphmag}}
\end{figure*}

\subsection{Divergence Cleaning of Magnetic Fields}
\label{sec:divb}

Both SPH and MFM are not able to strictly maintain {\em exactly} solenoidal magnetic fields naturally, and thus must employ cleaning schemes to keep their divergences minimal. We try to quantify the efficacy of this procedure in this section before showing our main results. 

We define the dimensionless divergence of magnetic fields as
\begin{equation}
  divB=\frac{h\vert \bm{\nabla}\cdot\bm{B}\vert}{B}.
\end{equation}
In our unstratified box simulations of section \ref{sec:unstrat}, the $divB$ diagnostic is smaller than $10^{-3}$ at the location of most fluid elements in MFM. Divergence control in the stratified shearing box MRI is more challenging, however. We run the fiducial model to compare the level of non-zero divergence in MFM with the CG cleaning \citep{Hopkins2015b, Hopkins2016a} and TSPH with the hyperbolic divergence cleaning \citep{Tricco2012}. In figure \ref{fig:divb} the hyperbolic cleaning keeps $divB\sim 0.1$ in TSPH while the CG cleaning keeps $divB$ two orders of magnitude lower in MFM. Large $divB$ only occurs at the vertical boundaries and in the weak field regions in MFM; the vertical boundaries are poorly resolved because MFM fluid elements,
which are built from particles, are fewer, but there the divergence should have negligible influence on the turbulent disk body since the correlation length of magnetic fields is smaller than \emph{H} \citep{Davis2010,Bai2013}.  We also stress that the maps shown in Figure \ref{fig:divb} are quite representative of the differences between SPH and MFM in our tests. 

Summarizing, the CG cleaning method in MFM significantly outperforms its competitors here, and we shall see how this is important in the following subsection. As a word
of caution, we note that, SPH methods that are recast at least
partially in a finite-volume formulation, such as Godunov-SPH \citep{Inutsuka2002} , might be amenable to implementations of the GC cleaning method. It would be interesting to explore the latter avenue in order to find out how much better an SPH method can 
perform once it is equipped with a superior divergence cleaning
scheme.


\subsection{Unphysical Behaviour in SPH Simulations}
\label{sec:sph}

\begin{figure}
  \plotone{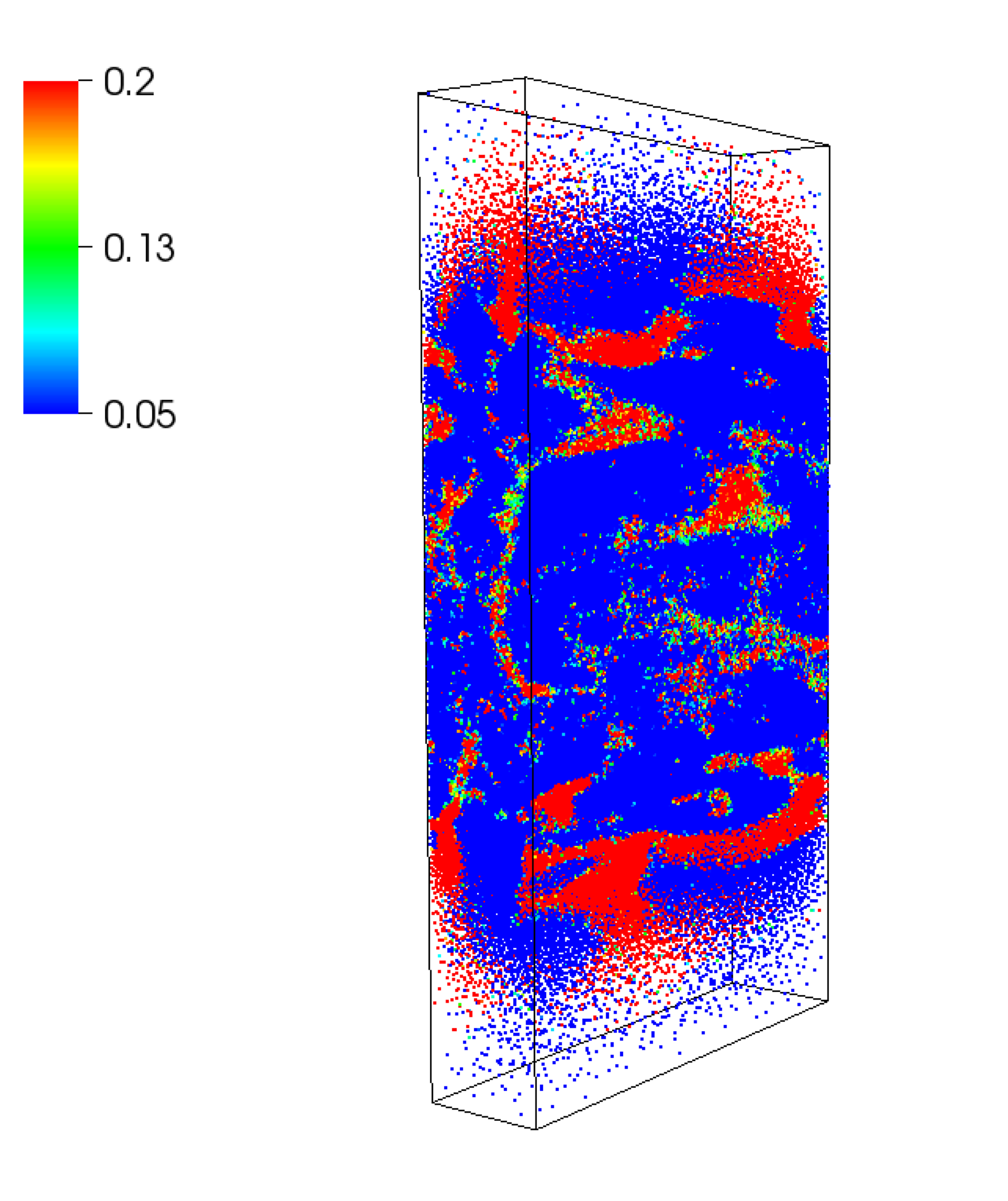}
  \caption{The artificial viscosity $\alpha_{sph}$ parameter in the PSPH run at $50\Omega^{-1}$ (see also figure \ref{fig:MF}). Relatively large artificial viscosity is triggered in the disk body where despite the fact that no shocks.\label{fig:av}}
\end{figure}

\begin{figure*}
  \epsscale{1.2}
  \plotone{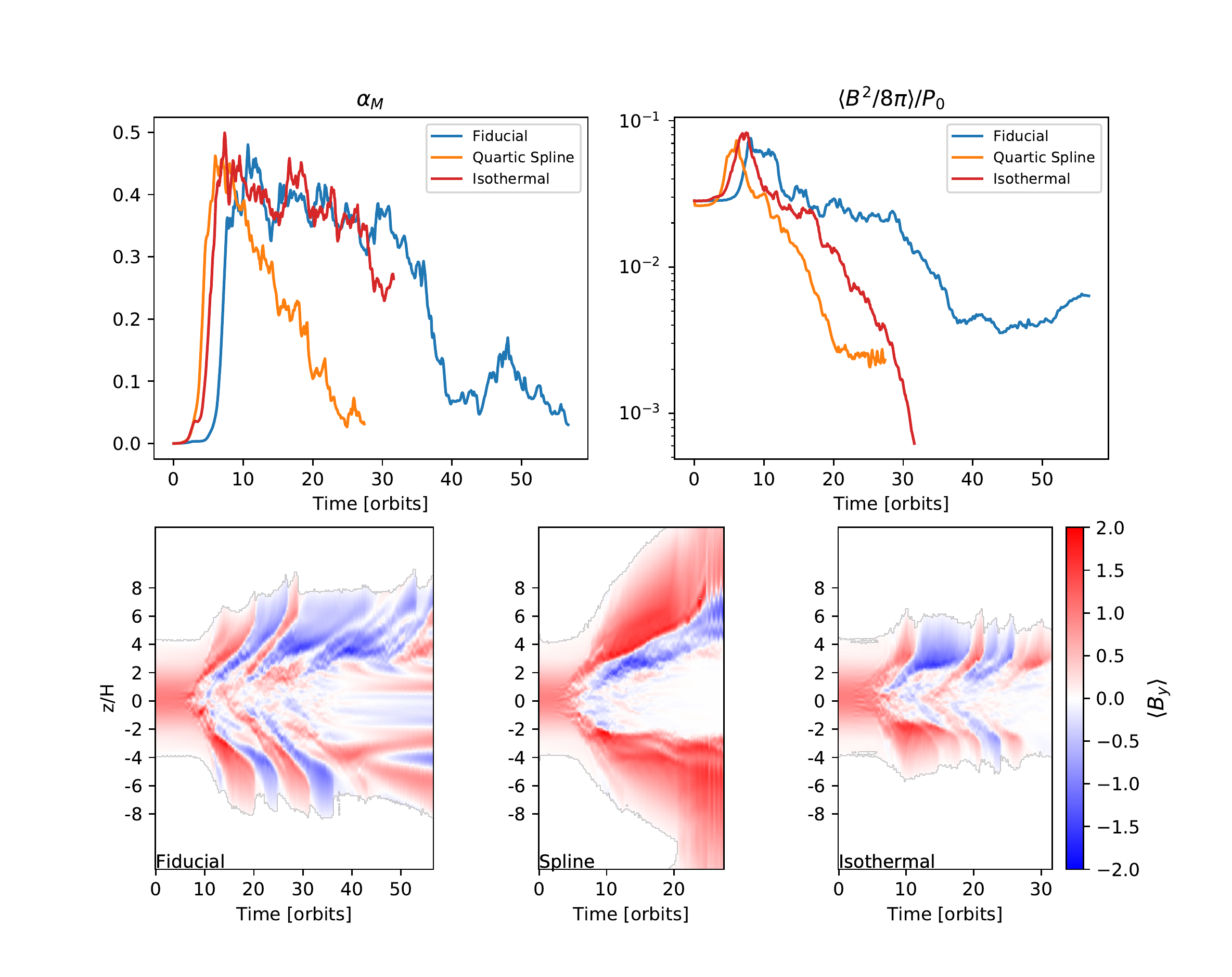}
  \caption{The evolution of the magnetic fields in the MFM stratified shearing box simulation with 1.5M elements and different setups. The upper panels are the time evolution of $\alpha_M$ and the magnetic energy as noted. The lower panels show the time evolution of the horizontally averaged azimuthal magnetic fields. In the fiducial run, both the saturated $\alpha_M$ and averaged azimuthal field pattern (butterfly diagram) agrees well with previous grid-code simulations \citep{Hawley2011,Simon2011} in the early 30 orbits. The fields decay later partially due to the expansion of the shearing box and thus decrease of the resolution (see figure \ref{fig:rho}). The simulation with the quartic spline kernel cannot reproduce the butterfly diagram due to numerical noise at the kernel scale, and the magnetic fields decay rapidly. The isothermal stratified shearing box with $\gamma=1.001$ doesn't expand vertically and thus maintains the resolution. However, the truncation error in the energy equation eventually leads to magnetic field dissipation.\label{fig:mfmmag}}
\end{figure*}

\begin{figure*}
  \plotone{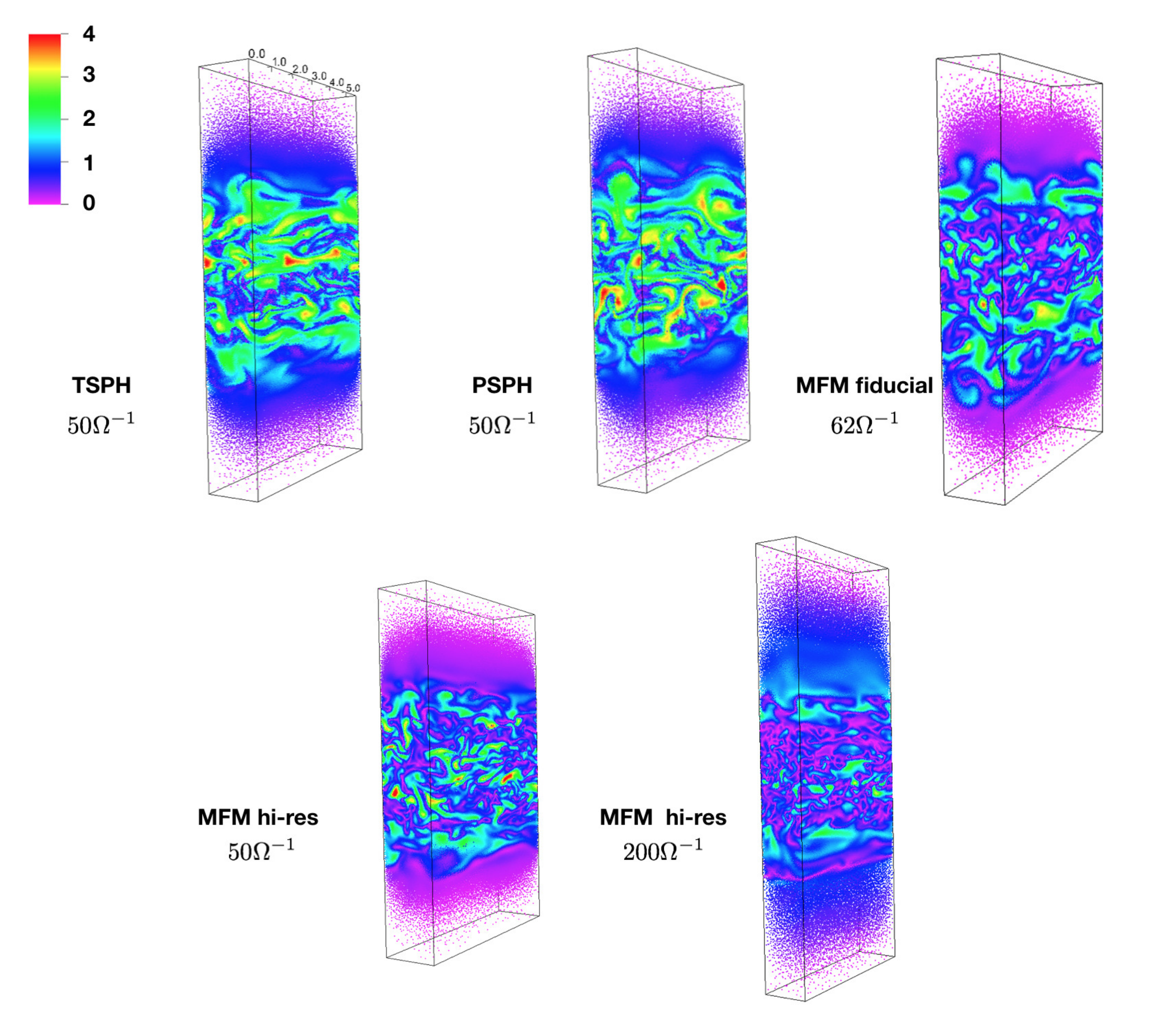}
  \caption{Fully developed MRI turbulence. Snapshots of magnetic field strength for a few stratified runs in table \ref{t:simulations} as labeled beside the panels (note the field strength is shown in Gauss and $\beta=1$ corresponds to magnetic field strength of 5 Gauss here). The upper panels are the fiducial 1.5M elements MFM simulation and its equivalent SPH simulations. The MFM snapshot is taken later than the SPH snapshots because the instability develops early in SPH due to stronger numeric noise (see figure \ref{fig:sphmag}, \ref{fig:mfmmag}). The snapshots are taken roughly when $\alpha_M$ reaches its maximum. Comparing to SPH simulations, MFM captures finer magnetic field structures and shows less noise in the fields. In the lower panels, the higher resolution stratified shearing box (3M elements) captures even finer structures.\label{fig:MF}}
\end{figure*}

We present three SPH simulations, two of which are run with the Wendland C4 kernel and adiabatic EOS but using different SPH formulations. In figure \ref{fig:sphmag} we plot $\alpha_M$ and the scaled magnetic energy versus time, in addition to space time diagrams of the horizontally averaged toroidal field. As is clear, the TSPH and PSPH simulations provide similar results. At first the MRI grows and expels the initial azimuthal fields to the disk corona where strong fields accumulate and are amplified (at about $\sim 60\Omega^{-1}$). Domains dominated by magnetic
energy propagate from the corona to the disk midplane and ultimately the entire box is dominated by strong growing azimuthal fields ($\beta\sim 1$). Note that $\alpha_M$ is negligible from some 10-20 orbits, indicative that the MRI is quenched. At the end of the simulation the magnetic fields are at equipartition with the gas pressure, and almost entirely azimuthal with no turbulent activity. Simultaneously the disk expands vertically as it becomes magnetically supported. As is obvious, these simulations bear little resemblance to previous grid-code stratified shearing box simulations, which report robust subsonic turbulence in the disk body \citep{Davis2010,Simon2011}. To test if the strong fields persist in  a more numerically dissipative setup, we reran the TSPH simulation with the quartic spline kernel (which has a noisier element distribution) and an isothermal EOS (see figure \ref{fig:mfmmag}). We find that this TSPH variant in fact damps the turbulence faster and reaches the $\beta \sim 1$ state earlier.

We should note, the result that SPH grows strong toroidal fields is not unique to our simulation setup or code. \citet{Dobbs2016}, using the SPHNG SPH code to simulate global galactic disk models, also reported unaccountable growth of magnetic fields. Likewise, similar behavior has also been seen in MHD-SPH simulations of disk formation in tidal disruption events with the code PHANTOM (Bonnerot, private communication). Both of these codes also implement the hyperbolic cleaning method from \citet{Tricco2012} in SPH. \citet{2015JCoPh.282..148S} presents a detailed study that discusses the likely numerical issues: they consider 3D, global simulations of a differentially rotating disk with an initially pure-toroidal field, designed so the system is stable and should exhibit no field growth. Using more accurate (CT or vector-potential based) schemes they show that they recover this solution. But using SPH with similar hyperbolic divergence cleaning, they show that discretization error produces small radial field components, which couples to the rotational shear and amplifies this and in turn the toroidal field exponentially. They specifically show that the form of the SPH MHD induction equation leads (in essentially any internally-consistent SPH based cleaning scheme) to the divergence-cleaning {\em amplifying} the vertical field, instead of {\em damping} the radial field, in order to locally restore $\nabla\cdot{\bf B}=0$.

This demonstrates a few key ingredients that interact here: the particularly virulent form of this instability in SPH requires shear/differential rotation (either in global disk simulations or shearing boxes), non-zero radial, azimuthal, and vertical field components where there is a vertical gradient present that can offset the radial gradient (hence 3D, stratified simulations), and relatively-large $\nabla\cdot{\bf B}$ errors (note these are large here, with $divB\sim 0.01-0.1$). 

We should also note that it is possible to construct divergence-cleaning schemes such as that in \citet{Tricco2012} which are total-energy conserving. In highly-idealized test problems, this will serve to limit the non-linear magnitude of any erroneous magnetic field amplification. However, in a shearing box or global thin disk simulation, there is an essentially infinite source of energy from shear, so this does not ``rescue'' the simulations from excessive numerical dissipation.

More generally, it is well-known that, without {\em any} divergence-cleaning, the $\nabla\cdot {\bf B}$ errors are violently numerically unstable: magnetic monopoles grow explosively and the amplitude of ${\bf B}$ is correspondingly rapidly-amplified. It is also well-established that this artificial, explosive field growth can occur even with divergence-cleaning, if the cleaning is not sufficiently accurate, or if it acts ``too slowly'' to respond to the growth rate. For example, \citet{Mocz2016} showed that using just \citet{Powell1999}-type (considerably less-sophisticated) divergence cleaning in even ordered meshes produces large artificial magnetic field growth (on essentially the Courant timescale) and much larger magnetic field strength, in idealized tests compared to CT methods.

Regarding the damping of turbulence, we should of course note that SPH requires artificial viscosity and resistivity to capture MHD shocks \citep[and references therein]{Cullen2010,Tricco2013}. It is well known that SPH tends to over-damp subsonic turbulence due to imperfectly triggered artificial viscosity \citep{Bauer2012, Hopkins2015a, Deng2017b}. The GIZMO code applies an artificial viscosity switch similar to that described in \citet{Cullen2010} with $\alpha_{min}=0.05$ and $\alpha_{max}=2$ \citep[see ][Appendix F2 for details]{Hopkins2015a} to suppress unwanted artificial viscosity. This switch works most efficiently in regions away from shocks and may not be effective at regions with large velocity derivatives \citep{Deng2017a}. In our SPH simulations, relatively large artificial viscosity with $\alpha_{sph}>0.2$ is still triggered (see figure \ref{fig:av}). Artificial viscosity certainly helps the turbulence dissipate. With damped velocity fluctuations, the MRI and its parasitic modes \citep{Latter2009,Pessah2009} cannot grow efficiently. We therefore also explore what happens if we revert to the more dissipative artificial viscosity in GADGET2 \citep{Springel2005} and restart the TSPH simulation from $t=50\Omega^{-1}$ (see figure \ref{fig:MF}); the turbulence does decay faster (as expected) and the strong toroidal fields develop more quickly. Thus, as expected, the MRI turbulence damping owes significantly to the artificial viscosity.

Artificial resistivity dissipates magnetic fields, and a switch to minimize artificial resistivity away from shocks was developed by \citet{Tricco2013}. We here apply this artificial resistivity switch with $\alpha_{B,min}=0.005$ and $\alpha_{B,max}=0.1$. We choose this conservative $\alpha_{B,max}$ because the turbulence is subsonic \citep{Hopkins2015b}.
We have re-run the TSPH simulation in figure \ref{fig:sphmag} with $\alpha_{B,max}=1$ \citep[suggested by][]{Tricco2013} and obtain similar results. The numerical resistivity in SPH MHD is evidently different from that of Riemann solvers (see appendix \ref{sec:eta}).

\subsection{A Transient MRI dynamo in MFM Simulations}
\label{sec:mfm}

Our MFM simulations use the same initial conditions as those of the SPH simulations. We present three simulations. The fiducial model is run with the Wendland C4 kernel and adiabatic EOS. In addition, to test the effect of the kernel function and EOS, we run two simulations with the quartic spline kernel ($N_{ngb}=60$) and with an isothermal EOS.  In the isothermal run we solve the energy equation instead of dropping it as done in \citet{Stone2008}. To mimic the isothermal EOS we set $\gamma=1.001$ so that the thermal energy dominates the total energy and large truncation errors affect the accuracy of magnetic energy calculation. This can cause the fast dissipation of the magnetic fields.

In figure \ref{fig:mfmmag} various quantities are plotted as functions of time. We see here that the MRI grows faster and reaches the magnetic pressure maximum quicker in the simulation with the quartic spline kernel (aqua) compared with the other two simulations. The quartic spline kernel is more compact and has a larger quality factor, $Q^*$ than the Wendland C4 kernel (see section \ref{sec:resolution}). Although \citet{Dehnen2012} shows that the quartic spline kernel is superior to the traditional cubic spline kernel it is still vulnerable to pair instability, which introduces numerical noise in gradient estimation. It would appear that this noise provides a significant degree of numerical dissipation because we find that the MRI dies quickly after its initial spike. The isothermal and fiducial adiabatic runs are qualitatively similar: both can sustain MRI turbulence for a period of some 30-40 orbits before dying.

\begin{figure}[ht!]
\plotone{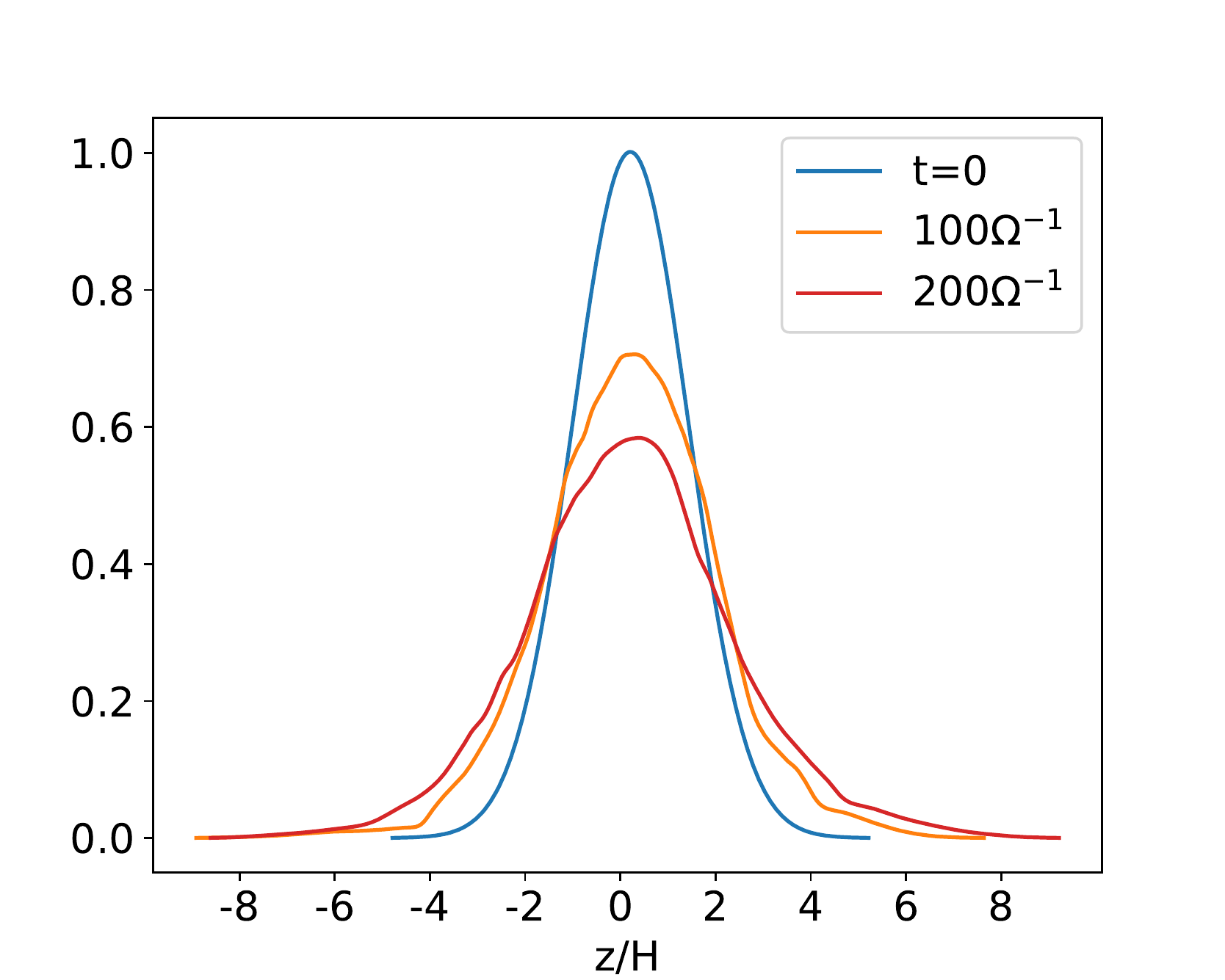}
\caption{The vertical density profile of the fiducial stratified shearing box MFM simulation at different time. The box expands vertically and at $200\Omega^{-1}$ the midplane density drops to 0.6 (the mean element separation becomes 1.2 times larger). \label{fig:rho}}
\end{figure}

In the first 30 orbits the fiducial model successfully reproduces the quasi-periodic ($\sim$10 orbits) butterfly pattern of the averaged azimuthal fields. Note, however, that the butterfly diagram becomes erratic at $\sim 200\Omega^{-1}$ as seen in other thermal MRI runs \citep{Gressel2013,Riols2018}. The saturated $\alpha_M\sim 0.4$ and $\langle B^2/8\pi \rangle/P_0 \sim 0.01$, however, are both in agreement with previous isothermal grid-codes' results \citep{Simon2011}. During the simulation, the box expands vertically due to accretion heating leading to a decrease of resolution in the disk body. In figure \ref{fig:rho}, the density at the disk midplane drops to 0.6 at $200\Omega^{-1}$ which corresponds to 1.2 times larger mean fluid element separation. The decrease in resolution certainly must affect the sustainability of the turbulence. We turn to higher resolution simulations to assess if our results can improve.

\subsection{High resolution MFM runs}
\label{sec:3M}

\begin{figure*}[ht!]
\epsscale{1.2}
\plotone{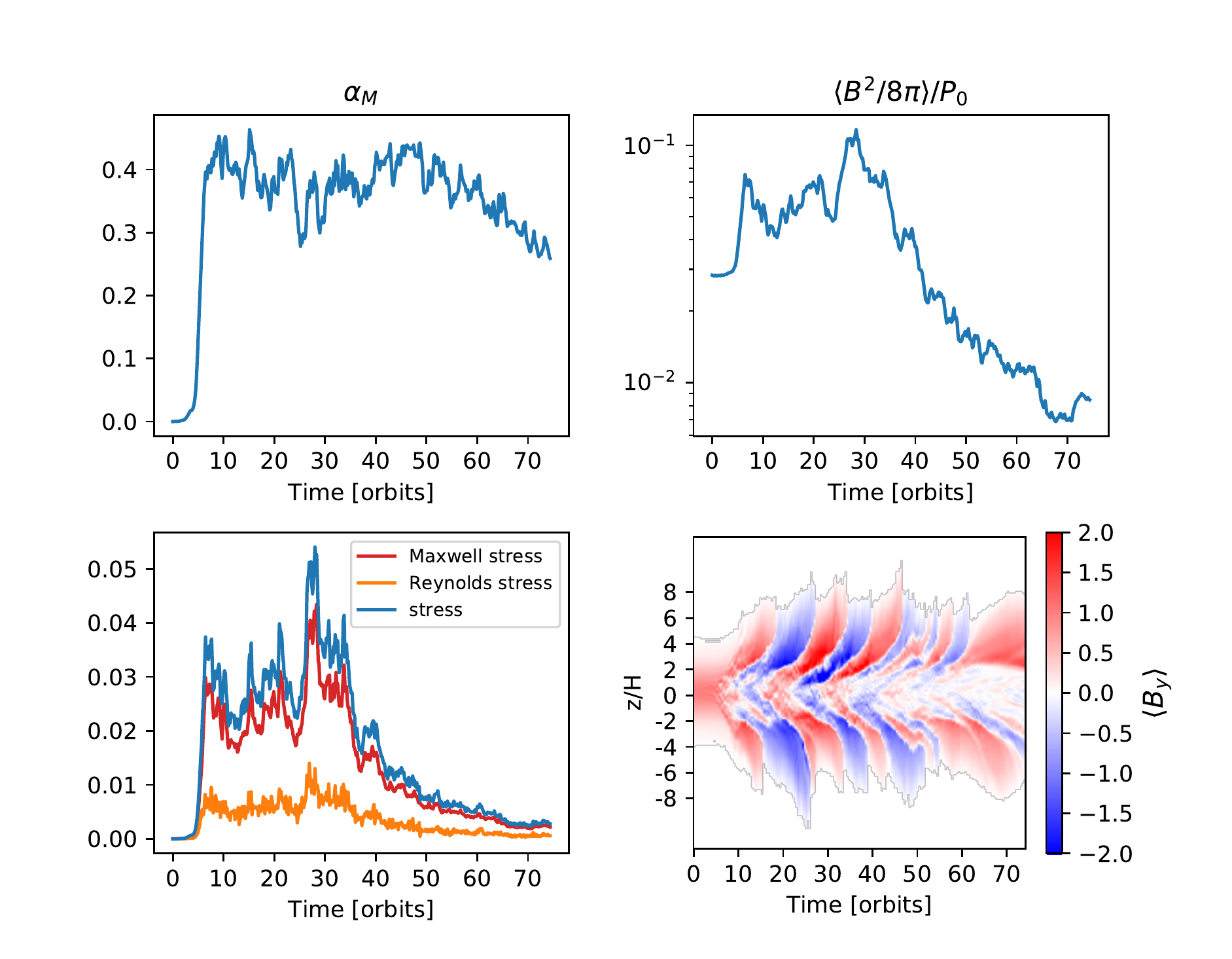}
\caption{The evolution of magnetic fields in the high resolution (3M elements) MFM stratified shearing box simulation. The saturated $\alpha_M$ is $\sim$0.4 with $\langle \beta\rangle \sim 100$. The Maxwell stress is roughly four times of the Reynolds stress as found in \citet{Hawley1996}. All stresses are normalized by $P_0$. The butterfly diagram becomes irregular at $\sim$50 orbits. \label{fig:3M}}
\end{figure*}

In order to maintain good resolution over the course of the
simulations, we avert expansion resulting from turbulent heat
transport and add an \emph{ad hoc} cooling term as in \citet{Noble2010, Parkin2013},
\begin{equation}\label{eq:cool}
  \frac{du_{cool}}{dt}=-\frac{u-u_{init}}{\tau_{cool}}
\end{equation}
where $\tau_{cool}=2\pi/\Omega$, and $u_{init}$ is the initial specific internal energy constant. This fast cooling maintains the disk scale height nearly constant, thus preserving
the initial resolution across the disk. In addition we increase the number of elements to 3 million which results in $\langle Q_y \rangle \sim 30, \langle Q_z \rangle \sim 10$ in the turbulent state.

In figure \ref{fig:3M} various flow properties are plotted. The most important result is that the MRI turbulence is sustained for a longer time (as it should if the method is converging properly). During this phase the main flow diagnostics are in good agreement with those of grid code runs: $\alpha_M \sim 0.4$, the averaged magnetic energy is a few percent of the gas pressure, and the Maxwell stress about 4 times larger than the Reynolds stress \citep{Hawley2011}.
In figure \ref{fig:3M}, the butterfly diagram is reproduced but after $300\Omega^{-1}$ the pattern becomes erratic. Comparing to the 1.5M elements MFM fiducial model in figure \ref{fig:mfmmag}, finer magnetic field structures are captured (see figure \ref{fig:MF}) and the butterfly diagram/dynamo is better resolved. We note that even this simple cooling can introduce additional numerical noise at the kernel scale \citep{Rice2014}. However, we cannot afford higher resolution,
or to run longer simulations, with this set-up (see section \ref{sec:hours}).

\section{Discussion}
\label{sec:dis}

\subsection{Computational Cost and Possible Applications}
\label{sec:hours}


In addition to robustness of numerical results, another worthy metric of comparison between codes is their computational cost to
carry out a comparable calculation. We run a setup equivalent to our 3M elements local stratified
simulation with ATHENA, and compare the computational cost directly using a fixed timestep in both. 
We found that the MFM simulation is $>100$ times more 
computationally expensive than an equivalent run with 32 cells per scale height using ATHENA with the orbital advection 
method \citep{Masset2000,Stone2010} for optimization. The lower computationally efficiency of MFM has nothing to do with
the hydro solver, rather
owes to the neighbor ``search tree'' which needs to  be updated constantly and walked to find 
neighbors and re-build the domain (because it allows for arbitrary particle re-configuration between timesteps). 
Of course, in simulations where particle order is not dramatically changing, and the only forces are local, 
we could in principle save considerable computational expense by simply storing the interacting neighbor lists and 
re-building the domain less often. Furthermore, there is room to improve significantly the neighbor search algorithm
on modern massively parallel architectures coupled with accelerators, as it is being currently investigated for
a range of particle-based codes \citep{Guerrera2018}.

Nevertheless, such a difference in performance is highly problem-dependent. The tree algorithm can be efficiently exploited -- and the difference in performance is dramatically mitigated -- if other physics which involves non-local forces is calculated. A prime example of the latter is self-gravity. The tree-based gravity solver
coupled with MFM, indeed, is, generally speaking, both faster and more accurate compared to traditional gravity solvers coupled with grid-based codes, which is
the reason why particle-based methods with tree-based gravity have been since long very competitive in comparison with grid-based methods in the modeling of 
self-gravitating protoplanetary disks \citep[see e.g.][]{Mayer2008}. Furthermore, addressing self-gravitating disks ultimately requires global calculations \citep{Durisen2007}. This is by itself a natural regime for mesh-free codes since one of their major goals is to enable adaptive resolution on global problems, more akin to adaptive-mesh-refinement codes, which have similar computational and memory cost. 

The most interesting applications of the mesh-free methods  studied here are thus not in idealized
MRI setups where accuracy of the MHD calculation over long timescales, absent other physical effects, is the prime objective.
Rather, these methods may be more promising for studies of turbulence in magnetized self-gravitating disks, especially the strong dynamo action reported by \citet{Riols2018}. This spiral wave dynamo is vigorous even with large magnetic resistivity and may be responsible for the primordial magnetic field amplification in galaxy formation \citep{Rieder2016,Rieder2017}, a field where adaptive resolution (either with Lagrangian or AMR-type codes) is essentially required. 
Likewise,  the methods described in this paper have considerable potential for applications in other areas of astrophysics where self-gravitating magnetized disks 
should be relevant, such as the central regions of massive protogalaxies where self-gravitating circumnuclear  gas disks could trigger the formation of supermassive black holes \citep{Regan2009,Choi2013,Mayer2010,Mayer2015} or the outer regions of accretion disks around AGNs \citep{Rafikov2001}.
In the case of the protogalactic nuclei, in particular, adaptivity is necessary to capture a wide range of spatial and temporal scales. Furthermore, 
understanding the interplay between the stabilizing effect of magnetic pressure, turbulence and gas inflows governed by global self-gravitating modes might
be the key to understand whether a monolithic central collapse into a supermassive star occurs, which later will turn into a massive black hole, as opposed to fragmentation into stars \citep{Latif2014}. The Riols \& Latter dynamo action might play
an important role in this latter case as it might reveal itself
as an important element to understand the process of angular 
momentum transport, and thus the evaluate better the possibility
of a central monolithic collapse.

\subsection{Other Lagrangian MHD methods}


We have restricted our study to just two classes of numerical methods, SPH and MFM (although we did consider a few ``variants'' of SPH). Furthermore, we only considered TSPH
and PSPH variants of the SPH method.
We should point out that caution is warranted in generalizing any of these results to other Lagrangian methods. Moving meshes or mesh-free finite-volume (MFV)-type methods with divergence-cleaning methods can arbitrarily ``smooth'' the mesh motion, decreasing the ``mesh deformation noise'' \citep{McNally2012,Munoz2014} and likely allowing for more accurate divergence cleaning simply because the mesh is deforming less rapidly and less irregularly (so e.g.\ smaller gradient errors can be ensured). As noted above, unstaggered CT schemes have now been developed \citep{Mocz2014, Mocz2016} for certain specific types of moving-mesh schemes, which can maintain $\nabla\cdot{\bf B}\approx 0$ at machine precision, so should perform more similarly to CT-grid schemes here, although the numerical noise/dissipation properties of moving-mesh codes (which determine the MRI damping) are often very different. 

Fundamentally distinct SPH MHD methods have also been developed. Although early attempts at implementing SPH MHD based on vector potentials did not allow reconnection \citep[e.g.][]{Rosswog2007}, newer hybrid methods that combine vector potentials with divergence-cleaning in the vector potential space appear to avoid exactly the runaway field amplification discussed here \citep[see][]{2015JCoPh.282..148S}. To our knowledge, however, these schemes have not yet been explored in a broader context or used for MRI simulations. 
Finally, in the final stages of the preparation of this paper we became aware that a simple variant of an SPH MHD solver based on the GDSPH method in the GASOLINE2 code \citep{Wadsley2017} is currently being tested in local MRI setups similar to those
described here (Robert Wissing et al, private communication).
In the latter, the Lorentz force is smoothed in the same way
as the hydro force, possibly helping to reduce numerical dissipation.

\section{Conclusions}
\label{sec:conclusion}

We presented the results of a series of MRI simulations with two meshless MHD methods, SPH and MFM, in both vertically unstratified and stratified boxes. Two variants of
SPH were considered, a ``vanilla" SPH method based on the density-energy formulation,
and PSPH (both as implemented in the GIZMO code).
The MRI, especially in its zero-net-flux configuration, is sensitive to numerical or physical dissipation which makes it challenging for any code to simulate, when physical diffusivities are omitted. This is true for both Eulerian and Lagrangian codes, and 
the results in the zero-net-flux case 
will always, to some extend, depend on the nature of their numerical dissipation. It is perhaps
then no surprise that the biggest discrepancies between the codes are observed for this magnetic
configuration. 

Our main findings can be summarized as:

1. The use of an appropriate kernel function which does not exhibit the pairing instability and allows a relatively large radius of compact support (e.g. Wendland C4) is crucial for maintaining element or mesh-generating-point order and for accurate gradient calculation. In MFM, this is directly akin to using a larger stencil to obtain more accurate, higher-order gradient estimators in traditional regular-grid codes. Although these kernels are less compact than the traditional spline kernels and tend to over-smooth fluid variables in SPH, they help to sustain the turbulence longer.

2. A stiff adiabatic EOS can help to control the noise in solving the energy equation where the truncation errors can be significant, because the magnetic energy is much smaller than the internal energy.

3. In unstratified shearing boxes with a net vertical field, MFM exhibits a similar error scaling in the linear growth MRI rates compared to the finite volume Eulerian code ATHENA. Both SPH and MFM can adequately simulate the ensuing turbulence, though the former is more diffusive and thus the MRI is closer to criticality. Two consequences of 
higher diffusivity in SPH are more vigorous channel bursts and very severe heating, as a consequence.

4. In unstratified shearing boxes with zero-net vertical field, SPH and MFM exhibit decaying turbulence at the (relatively low) resolution we are able to simulate here. It is possible this decay is linked to a very low numerical magnetic Prandtl number, but it is more likely that the numerical resistivity is simply too high at this resolution for sustaining the MRI.

5. In vertically stratified shearing box simulations, SPH MHD produces radically unphysical behaviour: turbulence dies out but strong toroidal fields continue to grow to equipartition with the gas pressure. This owes to non-trivial coupling of poorly-controlled magnetic field divergence, differential rotation/shear, and vertical stratification, at least in the most common SPH form of the induction and divergence-cleaning operators. 

6. In vertically stratified shearing boxes, high resolution MFM simulations produce results comparable to grid codes implementing the CT cleaning method. For several tens of orbits the classical MRI dynamo is captured, with its characteristic butterfly diagram. Nonetheless, the turbulence ultimately dies out after some 50 orbits, at the relatively low resolution studied as our ``baseline'' here. Going to higher resolution sustains the dynamo for longer, indicating that the decay is likely due to residual numerical resistivity.

We thank Matthew Bate, Daniel Price, Stephen Rosswog, Jim Stone, James Wadsley, Sijing Shen and Robert Wissing for useful discussions. We acknowledge support from the Swiss National Science Foundation via the NCCR PlanetS. Support for PFH was provided by an Alfred P. Sloan Research Fellowship, NSF Collaborative Research Grant \#1715847 and CAREER grant \#1455342, and NASA grants NNX15AT06G, JPL 1589742, 17-ATP17-0214. 

\software{GIZMO code \citep{Hopkins2015a}, VisIt\citep{HPV:VisIt}}
\bibliographystyle{aasjournal}
\bibliography{references}


\appendix

\section{Numerical viscosity}
\label{sec:nu}
\begin{figure}
  \epsscale{0.6}
  \plotone{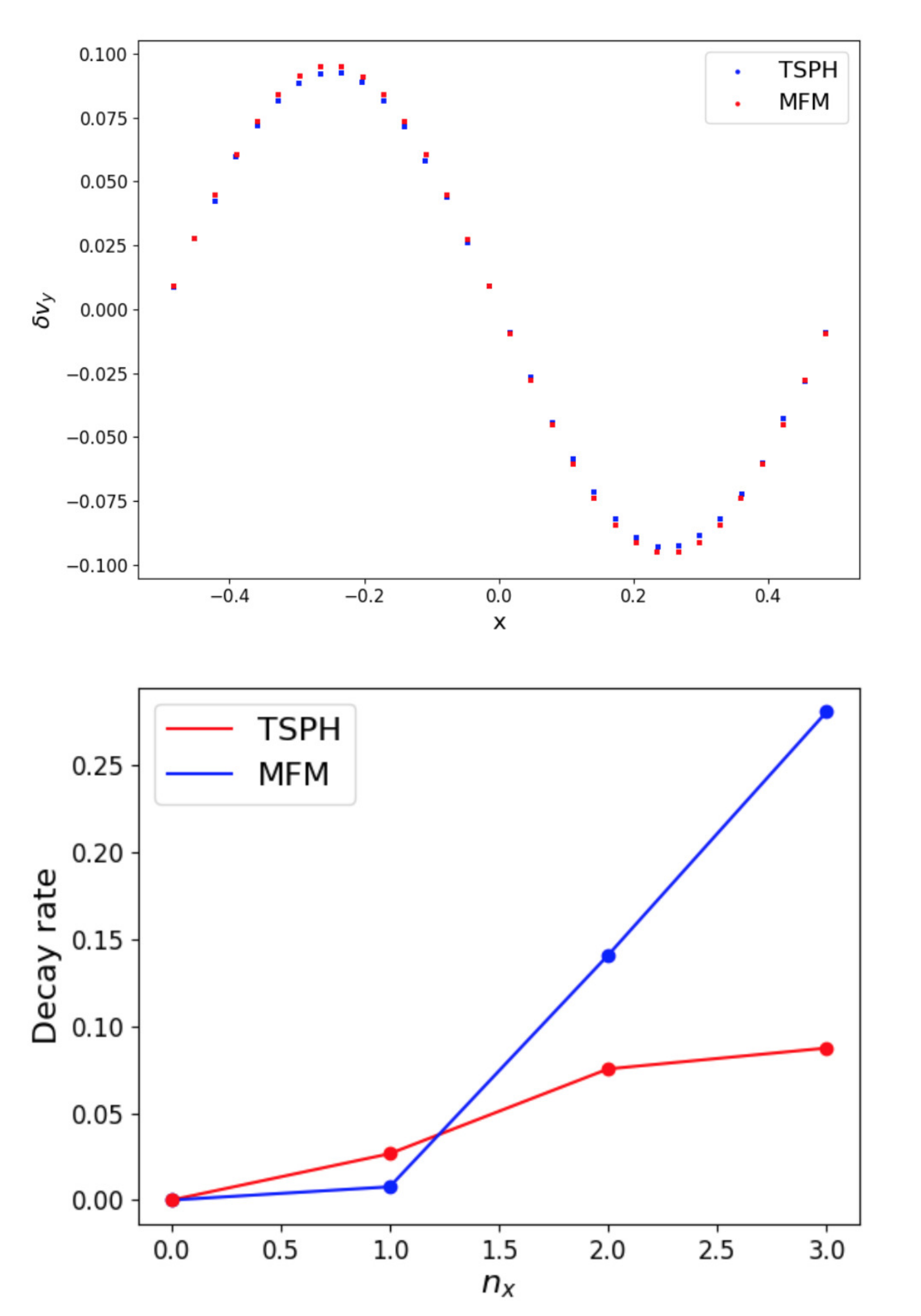}
  \caption{The velocity perturbation ($U=0.1, n_x=1$) decays slightly. In this test, the artificial viscosity is almost zero (in the Cullen \& Dehnen switch $\alpha_{sph}=\alpha_{min}=0.05$) and TSPH has a smaller numerical dissipation when the resolution is low. \label{fig:nu}} 
\end{figure}

To test the numerical viscosity we perturb the background shear flow by adding a radial dependent azimuthal velocity (a `zonal flow'), ie, $\delta{\bm{v}}=U sin(2\pi n_x x)\hat{\bm{y}}$. We adjust the internal energy (for the EOS used here, $\gamma=5/3$) so that the pressure perturbation is $\delta P=-\frac{\Omega U}{\pi n_x}cos(2\pi n_x x)$ and the initial setup is in equilibrium. Numerical viscosity will cause the perturbation to decay. By drawing an analogy to the Navier-Strokes equations (not necessarily true here), $U$ will decay at a rate $\nu_{num}k_x^{2}$, where $\nu_{num}$ is the effective numerical viscosity and $k_x=2\pi n_x$. We fit the decay of $\langle (\delta v_y)^2 \rangle$  to determine $\nu_{num}$ and the decay rate of $\langle (\delta v_y)^2 \rangle $ is shown in figure \ref{fig:nu}. We use a shearing box of size $H\times H\times H$ resolved by $32 \times 32 \times 32$ elements to carry out simulations with $U=0.1$ and $n_x=[1,2,3]$. In figure \ref{fig:nu}, the numerical viscosity in MFM is larger than TSPH when the resolution is low. However, MFM outperforms TSPH at 32 elements per wavelength which is in line with the channel flow growth rate test in section \ref{sec:channel}.

\section{Numerical resistivity}
\label{sec:eta}
\begin{figure}
  \epsscale{0.6}
  \plotone{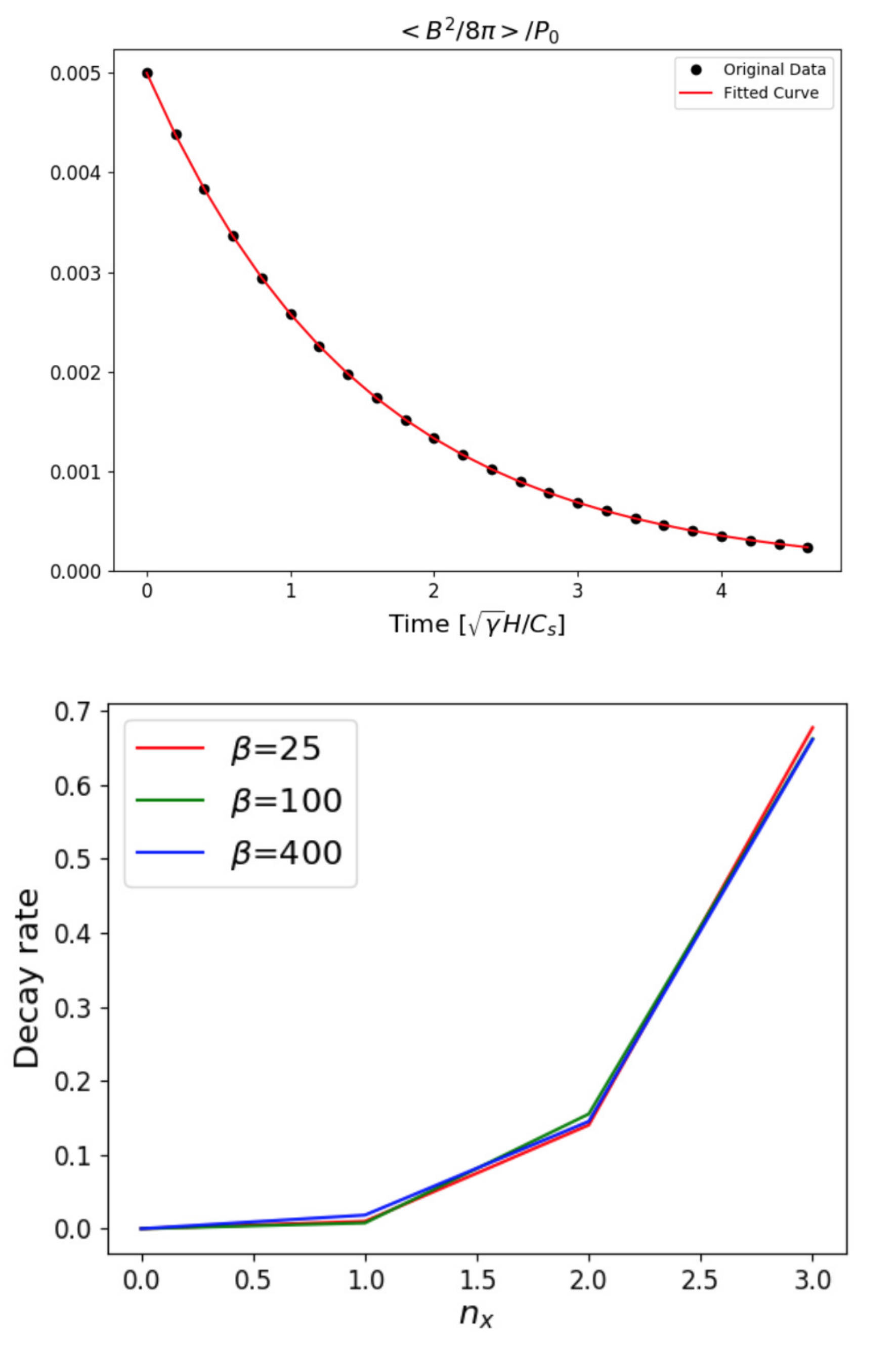}
  \caption{The magnetic energy decays exponentially in the test with $\beta =100, n_x=3$. We fit the curve to an exponential function to get the decay rate. The decay rate increase fast (fater than a parabola) as the resolution decreases, i.e., $n_x$ increases; it is almost independent of the magnetic field strength in our tested range.\label{fig:eta}}
\end{figure}

\begin{figure}
  \epsscale{0.6}
  \plotone{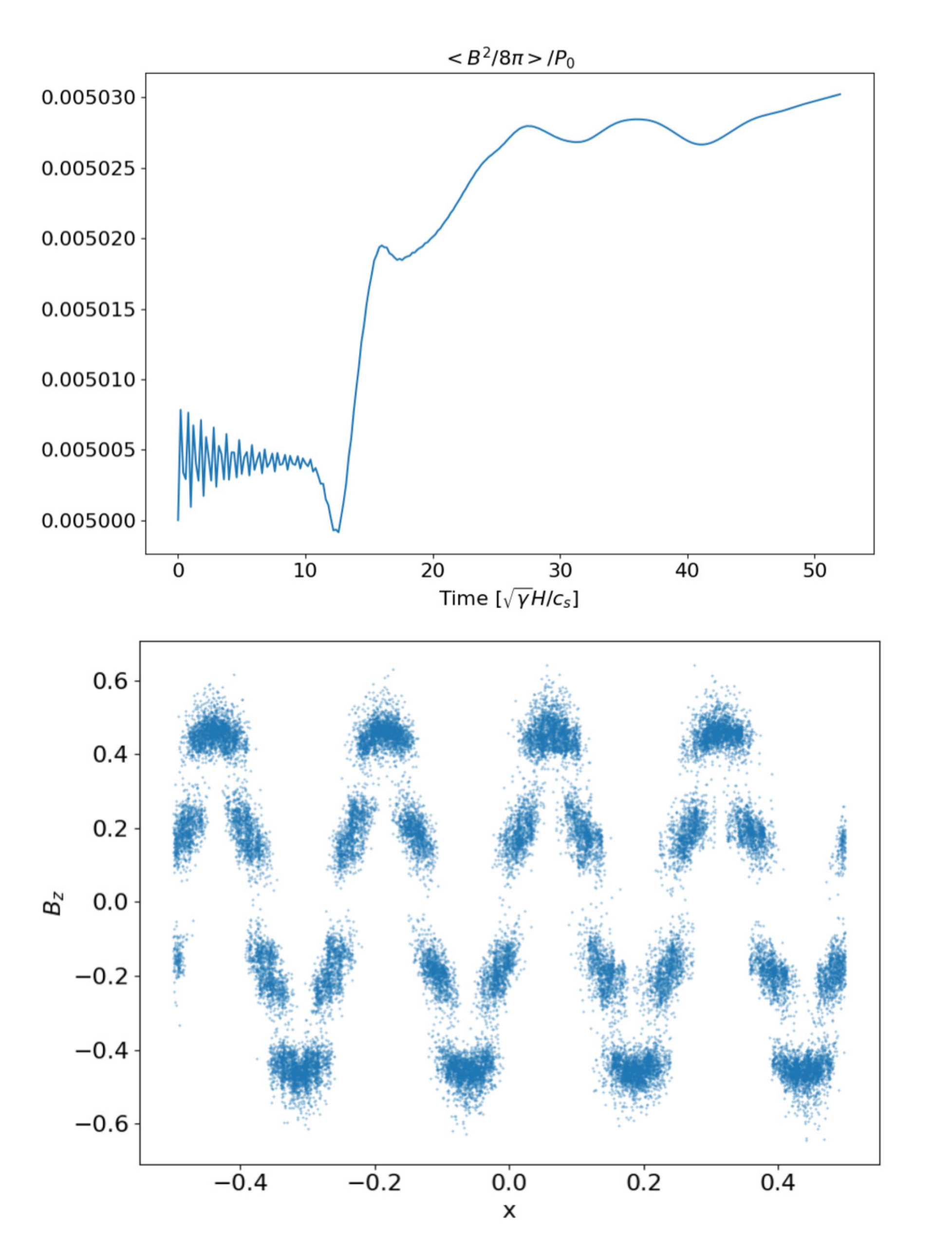}
  \caption{The magnetic energy doesn't decay in the TSPH test with $\beta =100, n_x=3$. But the magnetic field structure becomes noise at $50\sqrt{\gamma}H/c_s$\label{fig:spheta}} 
\end{figure}

Numerical dissipation can destroy magnetic fields. In a periodic box of size $H\times H \times H$ resolved by $32 \times 32 \times 32$ elements, we initial vertical magnetic fields and adjust the internal energy to set the box in pressure equilibrium. Here we set $\gamma=5/3$ but tests with an isothermal EOS behave similarly. The fields take the form of $\bm{B}=B_0 \hat{\bm{z}} sin(2\pi n_x x)$, where $B_0=\sqrt{8\pi P_0/\beta}$. In MFM simulations, the field structure decays due to numerical resistivity (see figure \ref{fig:eta}). $B_0$ should decay at a rate $\eta_{num}k_x^2$. The decay rate can be determined by fitting the decay of the averaged magnetic energy (decays twice as fast as $B_0$). We vary the wavelength ($n_x$) and field strength ($\beta$) to test how strong the numerical dissipation is. In this test, the dimensionless divergence of the magnetic fields is $\sim$0.0001 and we believe the dissipation is not caused by the non-zero divergence. However, in SPH simulations the magnetic energy doesn't decay even with $n_x=3$ (see figure \ref{fig:spheta}). The numerical noise from the artificial viscosity and resistivity break the perfect lattice. 

\end{document}